\documentclass[12pt,letterpaper]{article}

\usepackage{epsfig}

\begin{document}


\begin{center}
{\LARGE\bf Rationale for UV-filtered clover fermions}
\end{center}
\vspace{10pt}

\begin{center}
{\large\bf Stefano Capitani$\,{}^{a}$}
\hspace{0pt},\hspace{5pt}
{\large\bf Stephan D\"urr$\,{}^{b}$}
\hspace{5pt}{\large and}\hspace{5pt}
{\large\bf Christian Hoelbling$\,{}^{c}$}
\\[10pt]
${}^a\,$Institut f\"ur Physik, FB theoretische Physik,
Universit\"at Graz, A-8010 Graz, Austria\\
${}^b\,$Institut f\"ur theoretische Physik, Universit\"at Bern,
Sidlerstr.\,5, CH-3012 Bern, Switzerland\\
${}^c\,$Bergische Universit\"at Wuppertal,
Gaussstr.\,20, D-42119 Wuppertal, Germany
\end{center}
\vspace{10pt}

\begin{abstract}
\noindent
We study the contributions $\Sigma_0$ and $\Sigma_1$, proportional to $a^0$ and
$a^1$, to the fermion self-energy in Wilson's formulation of lattice QCD with
UV-filtering in the fermion action.
We derive results for $m_\mathrm{crit}$ and the renormalization factors
$Z_S,Z_P,Z_V,Z_A$ to 1-loop order in perturbation theory for several filtering
recipes (APE, HYP, EXP, HEX), both with and without a clover term.
The perturbative series is much better behaved with filtering, in particular
tadpole resummation proves irrelevant.
Our non-perturbative data for $m_\mathrm{crit}$ and $Z_A/(Z_mZ_P)$ show that
the combination of filtering and clover improvement efficiently reduces the
amount of chiral
symmetry breaking -- we find residual masses $am_\mathrm{res}=O(10^{-2})$.
\end{abstract}
\vspace{10pt}


\newcommand{\pa}{\partial}
\newcommand{\pas}{\partial\!\!\!/}
\newcommand{\Dsl}{D\!\!\!\!/\,}
\newcommand{\Psl}{P\!\!\!\!/\;\!}
\newcommand{\psl}{p\hspace{-2mm}/}
\newcommand{\hqu}{\hbar}
\newcommand{\ovr}{\over}
\newcommand{\til}{\tilde}
\newcommand{\pri}{^\prime}
\renewcommand{\dag}{^\dagger}
\newcommand{\<}{\langle}
\renewcommand{\>}{\rangle}
\newcommand{\gaf}{\gamma_5}
\newcommand{\lap}{\triangle}
\newcommand{\dal}{{\sqcap\!\!\!\!\sqcup}}
\newcommand{\trc}{\mathrm{tr}}
\newcommand{\Mpi}{M_\pi}
\newcommand{\Fpi}{F_\pi}

\newcommand{\al}{\alpha}
\newcommand{\be}{\beta}
\newcommand{\ga}{\gamma}
\newcommand{\de}{\delta}
\newcommand{\ep}{\epsilon}
\newcommand{\ve}{\varepsilon}
\newcommand{\ze}{\zeta}
\newcommand{\et}{\eta}
\renewcommand{\th}{\theta}
\newcommand{\vt}{\vartheta}
\newcommand{\io}{\iota}
\newcommand{\ka}{\kappa}
\newcommand{\la}{\lambda}
\newcommand{\rh}{\rho}
\newcommand{\vr}{\varrho}
\newcommand{\si}{\sigma}
\newcommand{\ta}{\tau}
\newcommand{\ph}{\phi}
\newcommand{\vp}{\varphi}
\newcommand{\ch}{\chi}
\newcommand{\ps}{\psi}
\newcommand{\om}{\omega}

\newcommand{\psb}{\overline{\psi}}
\newcommand{\etb}{\overline{\eta}}
\newcommand{\psd}{\psi^{\dagger}}
\newcommand{\etd}{\eta^{\dagger}}
\newcommand{\kh}{{\hat k}}
\newcommand{\qh}{{\hat q}}
\newcommand{\kb}{{\bar k}}
\newcommand{\qb}{{\bar q}}

\newcommand{\bdm}{\begin{displaymath}}
\newcommand{\edm}{\end{displaymath}}
\newcommand{\bea}{\begin{eqnarray}}
\newcommand{\eea}{\end{eqnarray}}
\newcommand{\beq}{\begin{equation}}
\newcommand{\eeq}{\end{equation}}

\newcommand{\mr}{\mathrm}
\newcommand{\mb}{\mathbf}
\newcommand{\Nf}{{N_{\!f}}}
\newcommand{\Nc}{{N_{\!c}}}
\newcommand{\ri}{\mr{i}}
\newcommand{\DW}{D_\mr{W}}
\newcommand{\DN}{D_\mr{N}}
\newcommand{\MeV}{\,\mr{MeV}}
\newcommand{\GeV}{\,\mr{GeV}}
\newcommand{\fm}{\,\mr{fm}}
\newcommand{\MSB}{\overline{\mr{MS}}}

\def\twocolcc#1{\multicolumn{2}{|c|}{#1}}

\hyphenation{topo-lo-gi-cal simu-la-tion theo-re-ti-cal mini-mum con-tinu-um}


\section{Introduction}


The Wilson formulation of lattice QCD breaks the chiral symmetry among the
light flavors \cite{Wilson:1975id,Wilson:1974sk}.
Accordingly, Wilson fermions undergo an additive (and multiplicative) mass
renormalization.
While this is not a problem in principle --~the explicit breaking disappears
if the lattice spacing $a$ is sent to zero \cite{Karsten:1980wd}~-- it
entails a number of complications in numerical work based on this formulation.
There are several strategies how the additive mass renormalization might be
reduced.
A popular choice, to augment the action by a clover term, has the merit of
reducing cut-off effects from $O(a)$ to $O(ag_0^2,...,a^2)$
\cite{Symanzik:1983dc,Sheikholeslami:1985ij,Heatlie:1990kg}.
Another possibility, referred to as UV-filtering, is to replace all covariant
derivatives in the fermion action by smeared descendents, as proposed in a
staggered context \cite{Blum:1996uf,Orginos:1999cr,Hasenfratz:2001hp}
and later applied to Wilson/clover fermions \cite{DeGrand:1998jq,
Bernard:1999kc,Stephenson:1999ns,Zanotti:2001yb,DeGrand:2002vu}.
We find that filtering indeed ameliorates important technical properties of the
Wilson operator, as does the clover term without filtering.
The real improvement, however, comes from combining the two.

With standard conventions the ($r\!=\!1$) Wilson operator takes the form
\beq
D_\mr{W}(x,y)={1\ovr2}\sum_\mu
\Big\{
(\ga_\mu-I) U_\mu(x)\de_{x+\hat\mu,y}-
(\ga_\mu+I) U_\mu\dag(x-\hat\mu)\de_{x-\hat\mu,y}
\Big\}
+{1\ovr2\ka}\de_{x,y}
\label{def_wils}
\eeq
where $I$ is the identity in spinor space.
The Sheikholeslami-Wohlert ``clover'' operator follows by adding a hermitean
contribution proportional to the gauge field strength
\cite{Sheikholeslami:1985ij}
\beq
D_\mr{SW}(x,y)=
D_\mr{W}(x,y)-{c_\mr{SW}\ovr2}\sum_{\mu<\nu}\si_{\mu\nu}F_{\mu\nu}\;\de_{x,y}
\label{def_clov}
\eeq
with $\si_{\mu\nu}\!=\!{\ri\ovr2}[\ga_\mu,\ga_\nu]$ and $F_{\mu\nu}$ the
hermitean ``clover-leaf'' operator.
In order to cancel the $O(a)$ contributions, the coefficient $c_\mr{SW}$ needs
to be properly tuned.
In perturbation theory one finds $c_\mr{SW}\!=\!1$ at the tree-level and a
correction proportional to the $n$-th power of $g_0^2$ at the $n$-loop level.
It is well known that for the standard ``thin link'' operator perturbation
theory shows rather bad convergence properties.
Therefore, the ALPHA collaboration has started a non-perturbative improvement
program \cite{ALPHA}.
Another approach is to resum the tadpole contributions \cite{Lepage:1992xa},
since they are quite sizable.
For filtered Wilson/clover quarks this might be different -- we elaborate on
``fat link'' perturbation theory \cite{Bernard:1999kc,Hasenfratz:2001tw,
DeGrand:2002va,DeGrand:2002vu}, and we compare these predictions to
(non-perturbative) data.
It turns out that filtered perturbation theory shows a much better convergence
behavior, but still, it does not describe the data very accurately.
The agreement is (at accessible couplings) much better than in the unfiltered
theory, but it is far from being completely satisfactory.
We find that the additive mass shift is two orders of magnitude smaller than
without filtering, and this is extremely useful in phenomenological studies.

The following two sections contain our perturbative results for UV-filtered
Wilson/clover fermions.
Sect.\,2 focuses on the additive mass shift with $1,2,3$ steps of
APE, HYP, EXP, HEX filtering and arbitrary improvement coefficient $c_\mr{SW}$.
Sect.\,3 contains our 1-loop results for the renormalization factors
$Z_S,Z_P,Z_V,Z_A$ with these filterings, a reminder how improved currents
are constructed, and a comment on tadpole resummation.
Sect.\,4 presents our non-perturbative data for the additive mass shift and
some renormalization factors, both with $c_\mr{SW}\!=\!0$ and
$c_\mr{SW}\!=\!1$.
Sect.\,5 contains our summary.
Details of ``fat-link'' perturbation theory, of an explicit mass shift
calculation and of the parameter dependence have been arranged in three
appendices.


\section{Additive mass shift with UV filtering in 1-loop PT}


In this paper we consider four types of filtering: APE, HYP, EXP, HEX.
The fist two are well known \cite{Albanese:1987ds,Hasenfratz:2001hp},
the third one has been named ``stout'' in \cite{Morningstar:2003gk},
and the fourth one is a straightforward application of the
hypercubic nesting trick on the latter (see App.\,A for details).
While on a technical level the smearing produces a smoothed gauge background,
it is in fact a different choice of the discretization of the covariant
derivative in the Dirac operator and therefore leads to an irrelevant change of
the fermionic action (provided the filtering recipe is unchanged when taking
the continuum limit).

In our analytical and numerical investigations we use the ``standard''
parameters
\beq
\al_\mr{std}^\mr{APE}\!=\!0.6\;,\qquad
\al_\mr{std}^\mr{EXP}\!=\!0.1
\label{std_APE_EXP}
\eeq
for APE and EXP smearing, and similarly the ``standard'' parameters
\beq
\al_\mr{std}^\mr{HYP}\!=\!(0.75,0.6,0.3)\;,\qquad
\al_\mr{std}^\mr{HEX}\!=\!(0.125,0.15,0.15)
\label{std_HYP_HEX}
\eeq
for HYP and HEX smearing.
The two values in (\ref{std_APE_EXP}) are related by giving an identical 1-loop
prediction for all quantities of interest (e.g.\ $-am_\mr{crit}$), and the same
statement holds for the hypercubically nested recipes (\ref{std_HYP_HEX}), see
App.\,A for details.
Accordingly, all perturbative tables with label ``APE'' will apply to EXP, too,
and ditto for a label ``HYP'' and the HEX recipe.

\begin{table}
\begin{center}
\begin{tabular}{|c|ccccc|}
\hline
{               }&thin\,link&  1\,APE  &  2\,APE  &  3\,APE  &  1\,HYP  \\
\hline
$c_\mr{SW}\!=\!0$& 51.43471 & 13.55850 &  7.18428 &  4.81189 &  6.97653 \\
$c_\mr{SW}\!=\!1$& 31.98644 &  4.90876 &  1.66435 &  0.77096 &  1.98381 \\
$c_\mr{SW}\!=\!2$&  1.10790 & -7.11767 & -5.48627 & -4.23049 & -4.41059 \\
\hline
\end{tabular}
\caption{Additive mass shift $S$ for ``thin link'' Wilson or clover fermions
and after APE or HYP filtering with standard parameters. The uncertainty is of
order one in the last digit quoted.}
\label{tab_addshift}
\end{center}
\end{table}

The additive mass shift is given by the self-energy $\Sigma_0$ via
[note that $am_\mr{crit}\!<\!0$ with (\ref{mquark_W})]
\beq
am_\mr{crit}=\Sigma_0=-{g_0^2\ovr16\pi^2}C_F S+O(g_0^4)
\label{def_amcrit}
\eeq
where $S$ is the quantity that is usually tabulated and $C_F\!=\!4/3$ for
$SU(3)$ gauge group.
Generalizing a standard calculation \cite{Capitani:2002mp} to ``fat-link''
perturbation theory (see App.\,A for a summary) one may work out
1-loop predictions for $S$ \cite{Bernard:1999kc,DeGrand:2002va}.
We have done this for arbitrary $c_\mr{SW}$.
From inspecting Tab.\,\ref{tab_addshift} one notices that $c_\mr{SW}\!=\!1$
alone reduces the additive mass shift by a factor $1.6$.
Filtering alone achieves a factor $3.8$ or $7.4$ with a single APE or HYP step,
respectively.
However, the combination reduces it by a factor $10.5$ or $26.0$, and hence
proves much more efficient than any one of the ingredients alone.
The tuned $c_\mr{SW}$ that would achieve zero mass shift is slightly above $2$
in the thin-link case, and slightly above $1$ in all cases with filtering.
This is the first indication that filtered $c_\mr{SW}\!=\!1$ clover fermions
break the chiral symmetry in a much milder way than filtered Wilson or
unfiltered clover fermions.
An important question is, of course, to which extent this is realized
non-perturbatively, and we shall address this issue in due course.


\section{Renormalization factors with UV filtering in 1-loop PT}


\subsection{Generic setup}

In general, the matrix elements of some operator $O_j^\mr{cont}(\mu)$ in the
continuum $\overline{\mr{MS}}$ scheme and its lattice counterparts
$O_k^\mr{latt}(a)$ are related by
\beq
\<.|O_j^\mr{cont}(\mu)|.\>=\sum_k Z_{jk}(a\mu) \<.|O_k^\mr{latt}(a)|.\>
\eeq
\beq
Z_{jk}(a\mu)
=\de_{jk}-{g_0^2\ovr16\pi^2}(\Delta_{jk}^\mr{latt}-\Delta_{jk}^\mr{cont})
=\de_{jk}-{g_0^2\ovr16\pi^2}C_F z_{jk}
\label{def_zX}
\eeq
with $C_F=4/3$ for $SU(3)$ gauge group.
Typically (e.g.\ for 4-fermion operators and a non-chiral action), $k$ runs
over other chiralities than $j$.
For 2-fermion operators, this mixing shows up at higher orders in an expansion
in the lattice spacing $a$, and packing it into the construction of improved
currents, one is left with the diagonal term in (\ref{def_zX}).
With our convention (which agrees with \cite{Capitani:2002mp}, but not with
\cite{DeGrand:2002vu}) a value $z_X\!>\!0$ signals $Z_X\!<\!1$.
Specifically (with $X\!=\!S,P,V,A$),
\bea
Z_S(a\mu)=1-{g_0^2\ovr4\pi^2}\Big[{z_S\ovr3}-\log(a^2\mu^2)\Big]\;,&\quad\quad&
Z_V=1-{g_0^2\ovr12\pi^2}z_V
\label{setup_ZS_ZV}
\\[2mm]
Z_P(a\mu)=1-{g_0^2\ovr4\pi^2}\Big[{z_P\ovr3}-\log(a^2\mu^2)\Big]\;,&\quad\quad&
Z_A=1-{g_0^2\ovr12\pi^2}z_A
\label{setup_ZP_ZA}
\eea
for the (pseudo-)scalar densities and the (axial-)vector currents,
with corrections of order $O(g_0^4)$ throughout.


\subsection{Results for $Z_S,Z_P,Z_V,Z_A$ for Wilson and clover fermions}

The same approach of combining FORM-based \cite{Vermaseren:2000nd} standard
perturbative procedures \cite{Capitani:2002mp} with ``fat-link'' perturbation
theory that has been used in the previous section for the additive mass shift,
allows one to work out the renormalization factors $Z_S,Z_P,Z_V,Z_A$ for
arbitrary $c_\mr{SW}$.

\begin{table}[t]
\begin{center}
\begin{tabular}{|c|ccccc|}
\hline
$c_\mr{SW}\!=\!0$&thin\,link&  1\,APE  &  2\,APE  &  3\,APE  &  1\,HYP  \\
\hline
$z_S$            & 12.95241 &  1.12593 & -1.53149 & -2.87223 & -1.78317 \\
$z_P$            & 22.59544 &  5.28288 &  1.07019 & -0.98025 &  0.51727 \\
$z_V$            & 20.61780 &  6.39810 &  3.62281 &  2.51381 &  3.38076 \\
$z_A$            & 15.79628 &  4.31963 &  2.32197 &  1.56782 &  2.23054 \\
\hline
$(z_P\!-\!z_S)/2$&  4.82152 &  2.07848 &  1.30084 &  0.94599 &  1.15022 \\
$z_V\!-\!z_A$    &  4.82152 &  2.07847 &  1.30084 &  0.94599 &  1.15022 \\
\hline
\end{tabular}
\hspace{2cm}
\caption{Coefficient $z_X$ in formula (\ref{def_zX}) for the renormalization
factor $Z_X$ with $X=S,P,V,A$ for $c_\mr{SW}\!=\!0$ Wilson fermions with APE or
HYP filtering with standard parameters.}
\label{tab_zX_wils}
\end{center}
\begin{center}
\begin{tabular}{|c|ccccc|}
\hline
$c_\mr{SW}\!=\!1$&thin\,link&  1\,APE  &  2\,APE  &  3\,APE  &  1\,HYP  \\
\hline
$z_S$            & 19.30995 &  4.11106 &  0.40606 & -1.43930 & -0.03678 \\
$z_P$            & 22.38259 &  4.80364 &  0.65185 & -1.33218 &  0.12845 \\
$z_V$            & 15.32907 &  3.31243 &  1.43934 &  0.82550 &  1.38517 \\
$z_A$            & 13.79274 &  2.96614 &  1.31645 &  0.77195 &  1.30255 \\
\hline
$(z_P\!-\!z_S)/2$&  1.53632 &  0.34629 &  0.12290 &  0.05356 &  0.08262 \\
$z_V\!-\!z_A$    &  1.53633 &  0.34629 &  0.12289 &  0.05355 &  0.08262 \\
\hline
\end{tabular}
\begin{tabular}{|c|}
\hline
1\,HYP\,\cite{DeGrand:2002vu}\\
\hline
  0.12 \\
 -0.04 \\
  1.38 \\
  1.30 \\
\hline
 -0.08 \\
  0.08 \\
\hline
\end{tabular}
\caption{Like Tab.\,\ref{tab_zX_wils}, but for $c_\mr{SW}\!=\!1$ clover
fermions. The last column has been adapted to our sign convention
[cf.\,(\ref{def_zX})] and suggests that there is a mislabeling in
Tab.\,III of Ref.\,\cite{DeGrand:2002vu}.}
\label{tab_zX_clov}
\end{center}
\begin{center}
\begin{tabular}{|c|ccccc|}
\hline
$c_\mr{SW}\!=\!2$&thin\,link&  1\,APE  &  2\,APE  &  3\,APE  &  1\,HYP  \\
\hline
$z_S$            & 22.90672 &  4.35133 &  0.06571 & -1.91937 & -0.43671 \\
$z_P$            & 26.24177 &  6.10928 &  1.39146 & -0.81914 &  0.80287 \\
$z_V$            &  8.95400 & -0.33664 & -1.07948 & -1.08366 & -0.89073 \\
$z_A$            &  7.28648 & -1.21561 & -1.74236 & -1.63378 & -1.51052 \\
\hline
$(z_P\!-\!z_S)/2$&  1.66753 &  0.87898 &  0.66288 &  0.55012 &  0.61979 \\
$z_V\!-\!z_A$    &  1.66752 &  0.87897 &  0.66288 &  0.55012 &  0.61979 \\
\hline
\end{tabular}
\hspace{2cm}
\caption{Like Tab.\,\ref{tab_zX_wils}, but for $c_\mr{SW}\!=\!2$. This nails
down the full polynomial dependence on $c_\mr{SW}$.}
\label{tab_zX_ovim}
\end{center}
\end{table}

Our results for $z_X$ with $X\!=\!S,P,V,A$ in the unimproved case
$c_\mr{SW}\!=\!0$ are summarized in Tab.\,\ref{tab_zX_wils}.
An important check is that $(z_P\!-\!z_S)/2$ and $z_V\!-\!z_A$ should coincide
\cite{Gupta:1996yt}.
The pertinent entries indicate that the integration routine yields at least
6 significant digits.

Our results for $z_X$ with $X\!=\!S,P,V,A$ in the improved case
$c_\mr{SW}\!=\!1$ are summarized in Tab.\,\ref{tab_zX_clov}.
Again we check the quality of the agreement between $(z_P\!-\!z_S)/2$ and
$z_V\!-\!z_A$.
Moreover, since these figures indicate the amount of chiral symmetry breaking
\cite{Gupta:1996yt}, it is instructive to compare the bottom lines of
Tab.\,\ref{tab_zX_wils} to those of Tab.\,\ref{tab_zX_clov}.
Improvement alone reduces $z_V\!-\!z_A$ by a factor $3.1$.
One step of APE or HYP filtering diminishes it by a factor $2.3$ or $4.2$,
respectively.
However, the combination of these recipes achieves a factor $13.9$ or $58.4$,
and hence proves much more efficient that any of the ingredients alone.
This is in line with the lesson learned from Tab.\,\ref{tab_addshift}.

\begin{figure}
\vspace{-2mm}
\begin{center}
\epsfig{file=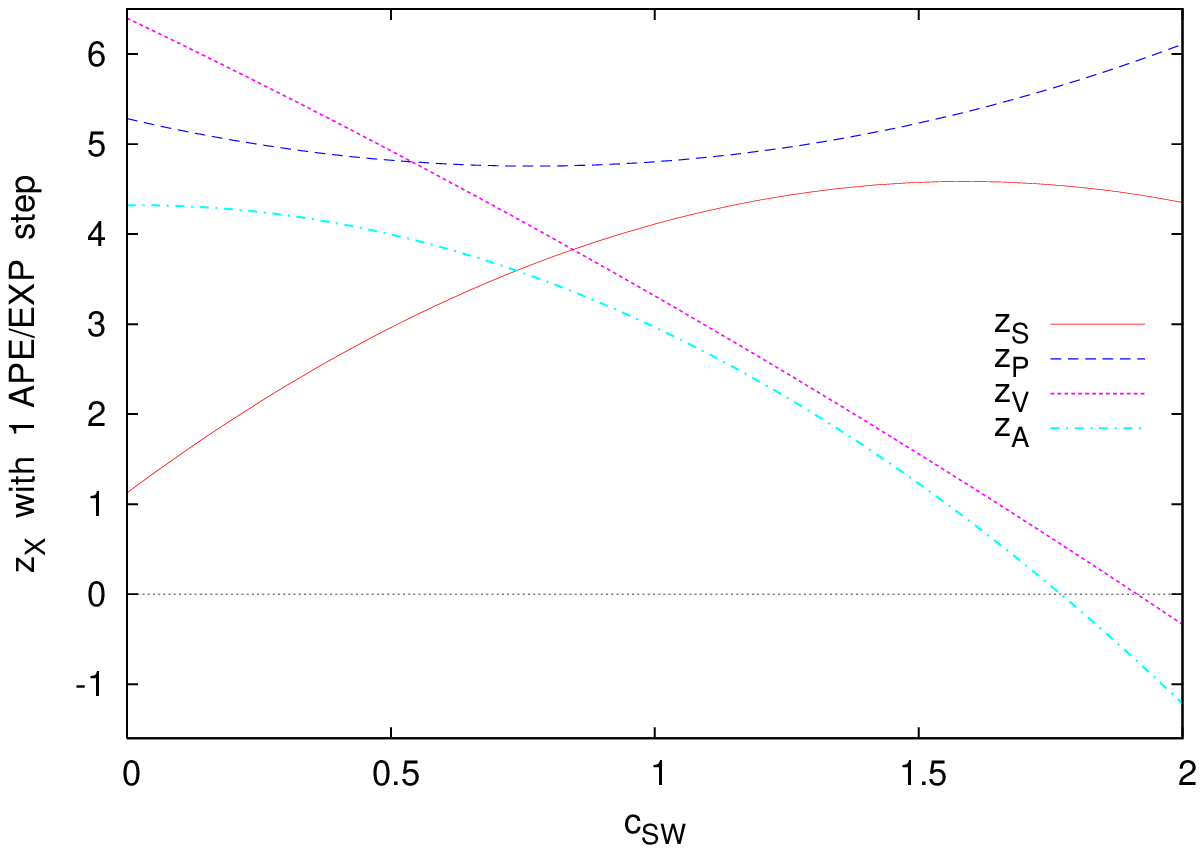,width=8.4cm}
\epsfig{file=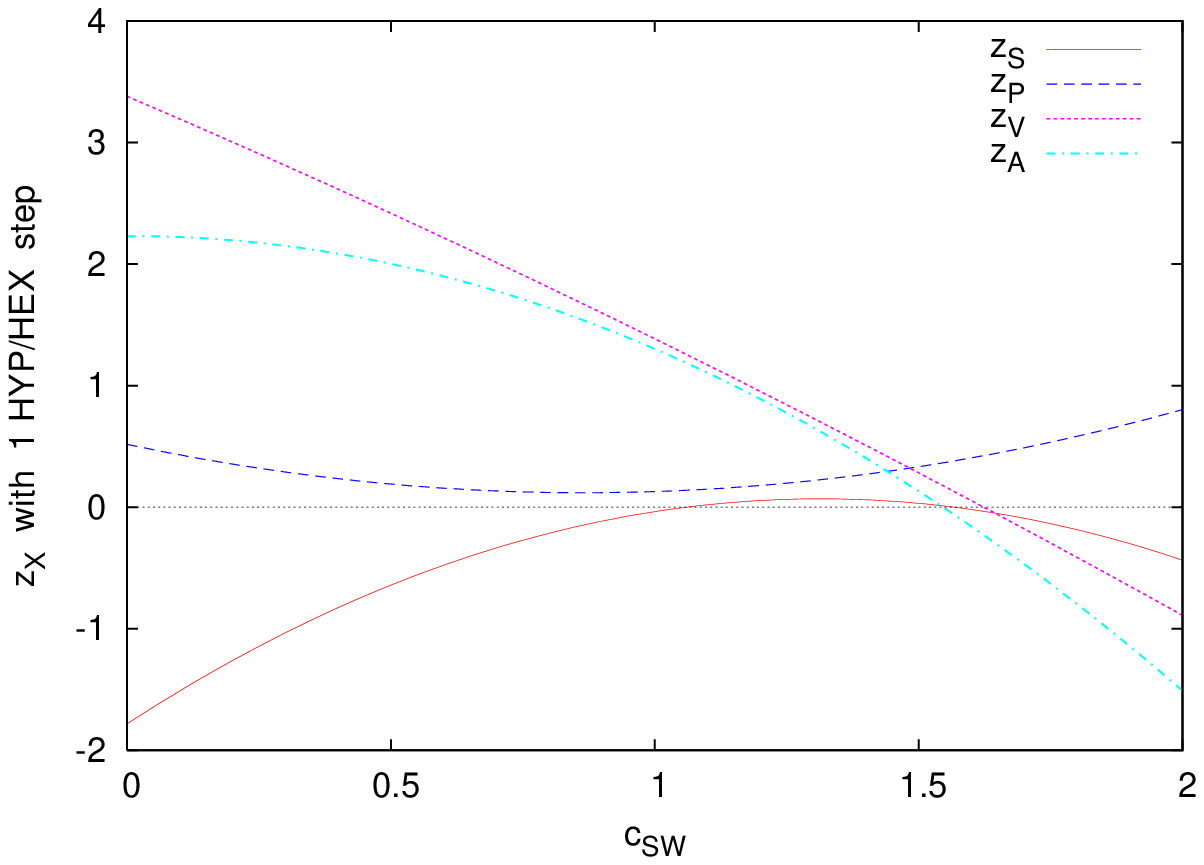,width=8.4cm}
\vspace{-2mm}
\caption{Finite pieces $z_{S,P,V,A}$ of the $Z_X$ for 1\,APE and 1\,HYP
fermions as a function of $c_\mr{SW}$.}
\vspace{-2mm}
\label{fig_zX_1HYP}
\end{center}
\end{figure}

Our results for $z_X$ in the case $c_\mr{SW}\!=\!2$ are shown in
Tab.\,\ref{tab_zX_ovim}.
Obviously, ``too much'' improvement deteriorates the chiral properties of the
action.
At 1-loop order all $z_X$ depend on $c_\mr{SW}$ through a quadratic polynomial,
hence Tabs.\,\ref{tab_zX_wils}\,-\,\ref{tab_zX_ovim} give them for arbitrary
values of the Sheikholeslami-Wohlert parameter.
For instance, for 1\,HYP (or 1\,HEX) step they imply
\bea
z_S&=& -1.78317 +2.81955 c_\mr{SW} -1.07316 c_\mr{SW}^2
\nonumber
\\
z_P&=& +0.51727 -0.92044 c_\mr{SW} +0.53162 c_\mr{SW}^2
\nonumber
\\
z_V&=& +3.38076 -1.85544 c_\mr{SW} -0.14015 c_\mr{SW}^2
\nonumber
\\
z_A&=& +2.23054 +0.01455 c_\mr{SW} -0.94254 c_\mr{SW}^2
\eea
and from the pertinent curves (see Fig.\,\ref{fig_zX_1HYP}) one learns two
lessons.
First, the point where the 1\,HYP action is most chiral (i.e.\ where
$z_P\!-\!z_S$ and $z_V\!-\!z_A$ are minimal) is near $c_\mr{SW}=1.1653$.
Second, near $c_\mr{SW}\!=\!1.5$ the four coefficients $z_{S,P,V,A}$ are
\emph{simultaneously} small.
By contrast, with less filtering (e.g.\ 1\,APE) the point of minimal chiral
symmetry breaking is further away from 1, and the four renormalization factors 
cannot be simultaneously close to 1.

Any strategy in which $c_\mr{SW}$ deviates, for large $\be$, from $1$ by a
polynomial in $g_0^2$ with vanishing constant part yields a theory with
$O(ag_0^2)$ cut-off effects.
Here we restrict ourselves to $c_\mr{SW}\!=\!1$.
Getting higher terms in the polynomial right reduces discretization effects to
$O(ag_0^4)$ or better, and non-perturbative improvement would realize $O(a^2)$.


\subsection{Construction of improved currents and densities}

At tree-level $Z_{S,P,V,A}\!=\!1$, and the improvement coefficients
are $c_\mr{SW}\!=\!1$, $b_{S,P,V,A}\!=\!1$, $b_m\!=\!-1/2$ and $c_{V,A}\!=\!0$.
Accordingly, in a tree-level $O(a)$ improved theory the currents read
\bea
(S_\mr{imp})^a    &=&(1+am_q)S^a
\nonumber
\\
(P_\mr{imp})^a    &=&(1+am_q)P^a
\nonumber
\\
(V_\mr{imp})_\mu^a&=&(1+am_q)V_\mu^a
\nonumber
\\
(A_\mr{imp})_\mu^a&=&(1+am_q)A_\mu^a
\eea
which is free of mixing effects, but it is well known that (at least in the
unfiltered case) this is not sufficient to be in the Symanzik $O(a^2)$ scaling
regime for accessible couplings.
Throughout, we use the flavor decomposition $X\!=\!X^a{\la^a\ovr2}$ with
$\la^a$ one of the Gell-Mann matrices ($a\!=\!1..8$).

At the 1-loop level and with $\Nf\!=\!0,\Nc\!=\!3$, renormalization factors in
the unfiltered theory%
\footnote{Throughout, we use $c_\mr{SW}$ to the previous order in quantities
which depend on it; these $Z_X$ are for $c_\mr{SW}=1$.}
take the form%
\footnote{With $\Nf\!>\!0$ they depend on $\til g_0^2=g_0^2(1+b_g am^\mr{W})$
with $b_g=0.012000(2)N_f$ and $m^\mr{W}$ given in (\ref{mquark_W})
\cite{ALPHA}.}
$Z_S\!=\!1\!-\!0.163042 g_0^2$,   
$Z_P\!=\!1\!-\!0.188986 g_0^2$,   
$Z_V\!=\!1\!-\!0.129430 g_0^2$,   
$Z_A\!=\!1\!-\!0.116458 g_0^2$,   
as follows from the first column of Tab.\,\ref{tab_zX_clov}.
Similarly
$c_\mr{SW}\!=\!1+0.2659g_0^2$     
\cite{Sheikholeslami:1985ij,Luscher:1996vw}, and
$b_S\!=1\!+\!0.1925 g_0^2$,       
$b_P\!=1\!+\!0.1531 g_0^2$,       
$b_V\!=1\!+\!0.1532 g_0^2$,       
$b_A\!=1\!+\!0.1522 g_0^2$,       
$b_m\!=\!-1/2\!-\!0.09623 g_0^2$, 
$c_V\!=\!-0.01633 g_0^2$,         
$c_A\!=\!-0.00757 g_0^2$,         
see \cite{ALPHA,Sint:1997jx,Taniguchi:1998pf,Aoki:2003sj} for details.
The main message is that most of the 1-loop corrections are large, since
$g_0^2\!\simeq\!1$.
With these expressions at hand, improved currents follow via
\bea
(S_\mr{imp})^a    &=&Z_S\,\til S^a  \,\quad,\qquad\,\til S^a=(1+b_S\,am_q)\,S^a
\nonumber
\\
(P_\mr{imp})^a    &=&Z_P\,\til P^a    \quad,\qquad  \til P^a=(1+b_P\,am_q)\,P^a
\nonumber
\\
(V_\mr{imp})_\mu^a&=&Z_V\,\til V_\mu^a\quad,\qquad  \til V_\mu^a=(1+b_V\,am_q)\,
[V_\mu^a+ac_V\bar\pa_\nu T_{\mu\nu}^a]
\nonumber
\\
(A_\mr{imp})_\mu^a&=&Z_A\,\til A_\mu^a\quad,\qquad  \til A_\mu^a=(1+b_A\,am_q)\,
[A_\mu^a+ac_A\bar\pa_\mu P^a]
\label{curr_impr}
\eea
where $\bar\pa_\mu\!=\!{1\ovr2}(\pa_\mu^{}\!+\!\pa_\mu^*)$ denotes the
forward-backward symmetric derivative.
Clearly, this is a complicated mixing pattern involving even
the tensor current.
Still, with perturbative coefficients it remains (in the unfiltered theory) a
challenge to reach those couplings where the Symanzik scaling with $O(a^2)$
cut-off effects sets in.
This is why (in the thin-link theory) a non-perturbative determination of the
renormalization constants and improvement coefficients is preferred
\cite{ALPHA}.

Our hope is that with filtering perturbative improvement at the 1-loop level is
a viable strategy.
An important check is how well the renormalized VWI quark mass and the
renormalized AWI quark mass coincide.
The (bare) Wilson or clover quark mass is defined as
\beq
m^\mr{W}=m_0-m_\mr{crit}
\qquad\mbox{where}\qquad
am_0={1\ovr2}\Big({1\ovr\ka}-{1\ovr\ka_\mr{tree}}\Big)
\;,\quad
am_\mr{crit}={1\ovr2}\Big({1\ovr\ka_\mr{crit}}-{1\ovr\ka_\mr{tree}}\Big)
\label{mquark_W}
\eeq
with $\ka_\mr{tree}\!=\!1/8$, and the (renormalized) VWI quark mass then
follows through
\beq
m^\mr{VWI}(\mu)=Z_m(a\mu)(1+b_m am^\mr{W})m^\mr{W}\;.
\label{mquark_VWI}
\eeq
The (bare) PCAC quark mass is defined through (for $A_\mu$ and $P$ built from
degenerate quarks)
\beq
m^\mr{PCAC}={1\ovr2}
{\<\bar\pa_\mu[A_\mu^a(x)+ac_A\bar\pa_\mu P^a]O^a(0)\> \ovr
 \<P^a(x)O^a(0)\>}
\label{mquark_PCAC}
\eeq
and the (renormalized) AWI quark mass then follows through
\beq
m^\mr{AWI}(\mu)={Z_A\ovr Z_P(a\mu)}\,
{1+b_A am^\mr{W}\ovr 1+b_P am^\mr{W}}\,m^\mr{PCAC}\;.
\label{mquark_AWI}
\eeq
In (\ref{mquark_VWI},~\ref{mquark_AWI}) the details of the conversion from
the specific cut-off scheme on the r.h.s.\ to the standard
$\overline{\mr{MS}}$-scheme on the l.h.s.\ are built into the renormalization
factors.
If we had $c_\mr{SW}$ and the $b_S,b_P,b_V,b_A,b_m,c_V,c_A$ at 1-loop level,
plus the $Z_S,Z_P,Z_V,Z_A$ at 2-loop level, a theory with $O(ag_0^4)$ cut-off
effects could be realized.
At the time, we lack the knowledge of any improvement coefficient at the
1-loop level (with filtering).
Accordingly, the following section is devoted to a preliminary test with
tree-level improvement coefficients and 1-loop renormalization factors.
Still, since the perturbative series converges so well, our hope is that this
test does not fail completely -- otherwise higher order corrections could
barely save the case.


\subsection{Irrelevance of tadpole resummation}

One of the attractive features of filtered Dirac operators is that 1-loop
renormalization factors and improvement coefficients are much closer to their
tree-level values, suggesting a better convergence pattern.
Obviously, a first guess says this is mostly due to the tadpole contribution
being much smaller than in the unfiltered theory.

\begin{table}
\begin{center}
\begin{tabular}{|c|cccccccc|}
\hline
  & $0.12$ & $0.24$ & $0.36$ & $0.48$ & $0.6$  & $0.72$ & $0.84$ & $0.96$ \\
\hline
1 &10.05384& 8.25363& 6.83240& 5.79017& 5.12693& 4.84269& 4.93744& 5.41118\\
2 & 8.50285& 6.32137& 5.11011& 4.45066& 4.08470& 3.91401& 4.00042& 4.56587\\
3 & 7.37921& 5.28658& 4.37748& 3.94939& 3.72277& 3.60854& 3.69503& 4.45440\\
4 & 6.55107& 4.68477& 4.00340& 3.70447& 3.54744& 3.46295& 3.55729& 4.69838\\
5 & 5.93054& 4.30964& 3.78626& 3.56346& 3.44535& 3.37845& 3.48725& 5.34886\\
\hline
\end{tabular}
\caption{Tadpole diagram in Feynman gauge [value to be multiplied with
$g_0^2C_F/(16\pi^2)$] in 1-loop ``fat-link'' perturbation theory. The
corresponding ``thin-link'' value is 12.233050.}
\label{tab_tadpole_fey}
\end{center}
\begin{center}
\begin{tabular}{|c|cccccccc|}
\hline
  & $0.12$ & $0.24$ & $0.36$ & $0.48$ & $0.6$  & $0.72$ & $0.84$ & $0.96$ \\
\hline
1 & 6.99558& 5.19536& 3.77414& 2.73191& 2.06867& 1.78443& 1.87918& 2.35292\\ 
2 & 5.44459& 3.26311& 2.05185& 1.39240& 1.02644& 0.85574& 0.94215& 1.50761\\
3 & 4.32095& 2.22832& 1.31922& 0.89113& 0.66450& 0.55028& 0.63677& 1.39614\\
4 & 3.49281& 1.62650& 0.94513& 0.64620& 0.48918& 0.40469& 0.49903& 1.64011\\
5 & 2.87228& 1.25138& 0.72799& 0.50519& 0.38709& 0.32019& 0.42898& 2.29060\\
\hline
\end{tabular}
\caption{Tadpole diagram in Landau gauge [value to be multiplied with
$g_0^2C_F/(16\pi^2)$] in 1-loop ``fat-link'' perturbation theory. The
corresponding ``thin-link'' value is 9.174788.}
\label{tab_tadpole_lan}
\end{center}
\end{table}

In Feynman gauge  the ``thin-link'' tadpole diagram with the value
$12.233050g_0^2C_F/(16\pi^2)$, which is responsible for many of the large
corrections in unfiltered perturbation theory \cite{Lepage:1992xa}, gets
reduced as detailed in Tab.\,\ref{tab_tadpole_fey} for a broad range of
$\al^\mr{APE}$ and $n_\mr{iter}$ parameters.
Note that these numbers hold for arbitrary $c_\mr{SW}$, since the dependence on
the Sheikholeslami-Wohlert parameter comes through quark-gluon vertices with an
odd number of gluons.

In Landau gauge the effect is even more pronounced, as shown in 
Tab.\,\ref{tab_tadpole_lan}.
Here, the ``thin link'' value is $9.174788g_0^2C_F/(16\pi^2)$, and a smearing
parameter $\al^\mr{APE}\!<\!\al_\mr{max}^\mr{APE}\!=\!0.75$ seems to be
beneficial (cf.\ App.\,A for details on $\al_\mr{max}^\mr{APE}$).
In this gauge the sunset diagram is rather small, regardless of the filtering
level.
We checked that, for the extreme choice
$(\al^\mr{APE},n_\mr{iter})\!=\!(0.45,10)$, we reproduce the result
$0.2597053g_0^2C_F/(16\pi^2)$ of \cite{Bernard:1999kc}.


From this observation it is plausible that tadpole improvement is not necessary
-- i.e.\ has barely an effect -- in fat-link perturbation theory.
This leaves us optimistic that the perturbative series might converge much
better for filtered actions.
The real issue is, of course, whether such perturbative predictions will agree
with non-perturbative data.


\section{Non-perturbative tests}


Here, we investigate how well a perturbative improvement program with 1-loop
renormalization factors and tree-level improvement coefficients works with
filtered Wilson/clover fermions.
Since no phenomenological insight is attempted, we work in the quenched theory.
We wish to cover a regime of couplings from $\be\!\simeq\!5.8$ to
$\be\!\simeq\!6.4$ with the Wilson (plaquette) action and we work in a fixed
physical volume as defined through the Sommer radius $r_0$ \cite{Necco:2001xg}.
The corresponding parameters (realizing $L/r_0\!=\!2.98$, and thus
$L\!\simeq\!1.49\fm$ if $r_0\!=\!0.5\fm$) are given in Tab.\,\ref{tab_params}.

\begin{table}
\begin{center}
\begin{tabular}{|c|ccccc|}
\hline
$\be$            & 5.846 & 6.000 & 6.136 &  6.260 & 6.373 \\
$L/a$            &   12  &   16  &   20  &    24  &   28  \\
$L/r_0$          & 2.979 & 2.981 & 2.983 &  2.981 & 2.979 \\
$a^{-1}\,[\GeV]$ & 1.590 & 2.118 & 2.646 &  3.177 & 3.709 \\
$n_\mr{conf}$    &   64  &   32  &   16  &     8  &    4  \\
\hline
\end{tabular}
\caption{Matched ($\be$, $L/a$) combinations to achieve $L/r_0\!=\!2.98$ as
accurately as possible, based on the interpolation formula of
\cite{Necco:2001xg}. $n_\mr{conf}$ is the number of configurations per
filtering and mass.}
\label{tab_params}
\end{center}
\end{table}

Technically, we produce a smeared copy of the actual gauge field, and evaluate
the fermion action on that smoothed background.
This differs from the approach taken in \cite{Zanotti:2001yb}, since our entire
$D_\mr{W}$ in (\ref{def_clov}) is constructed from smoothed links.
See App.\,B for details.


\subsection{Data for $m_\mr{crit}$, $\til Z_A$ with APE/HYP/EXP/HEX filtering}

\begin{figure}
\vspace{-1mm}
\epsfig{file=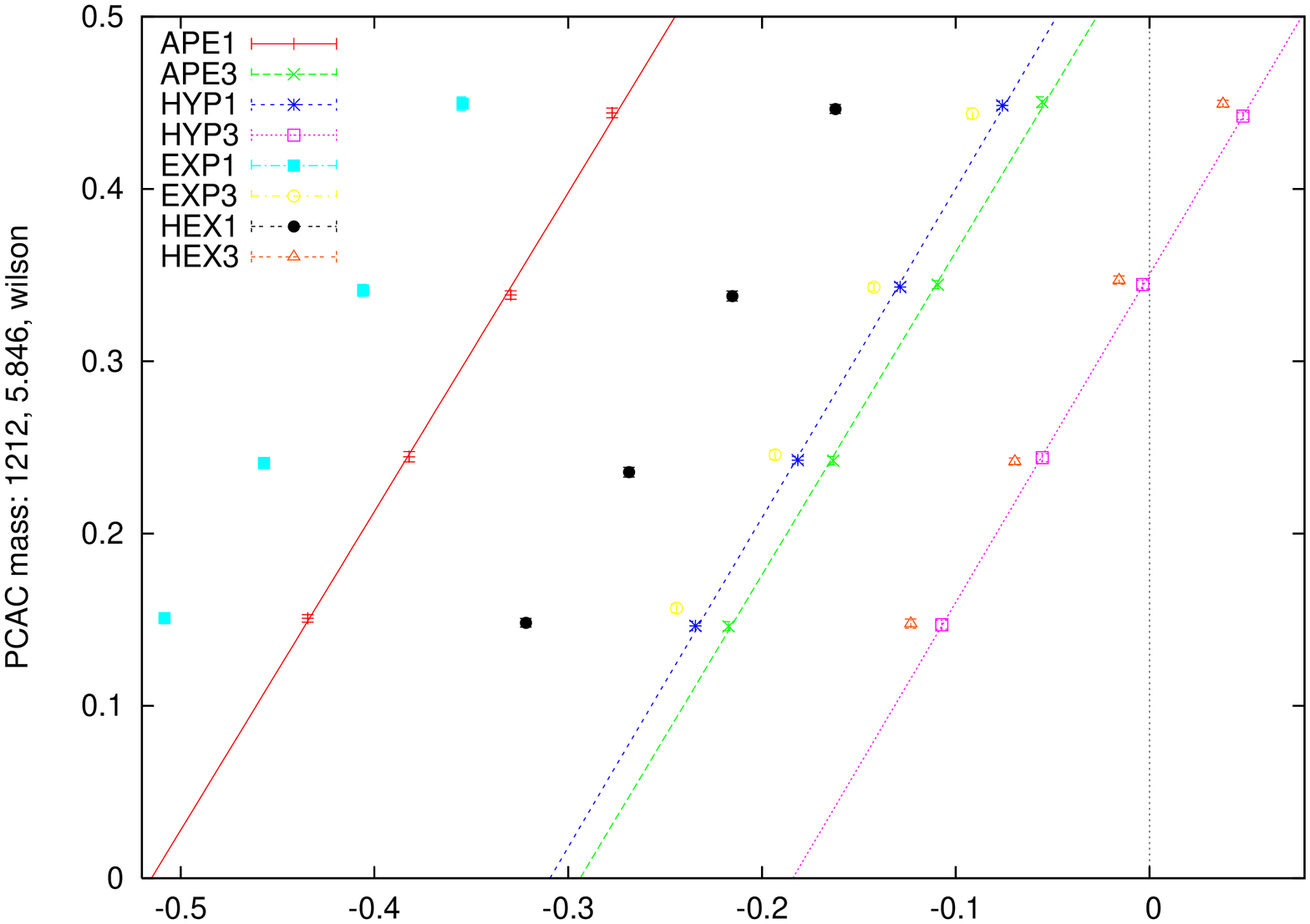,height=85mm,angle=-90}
\epsfig{file=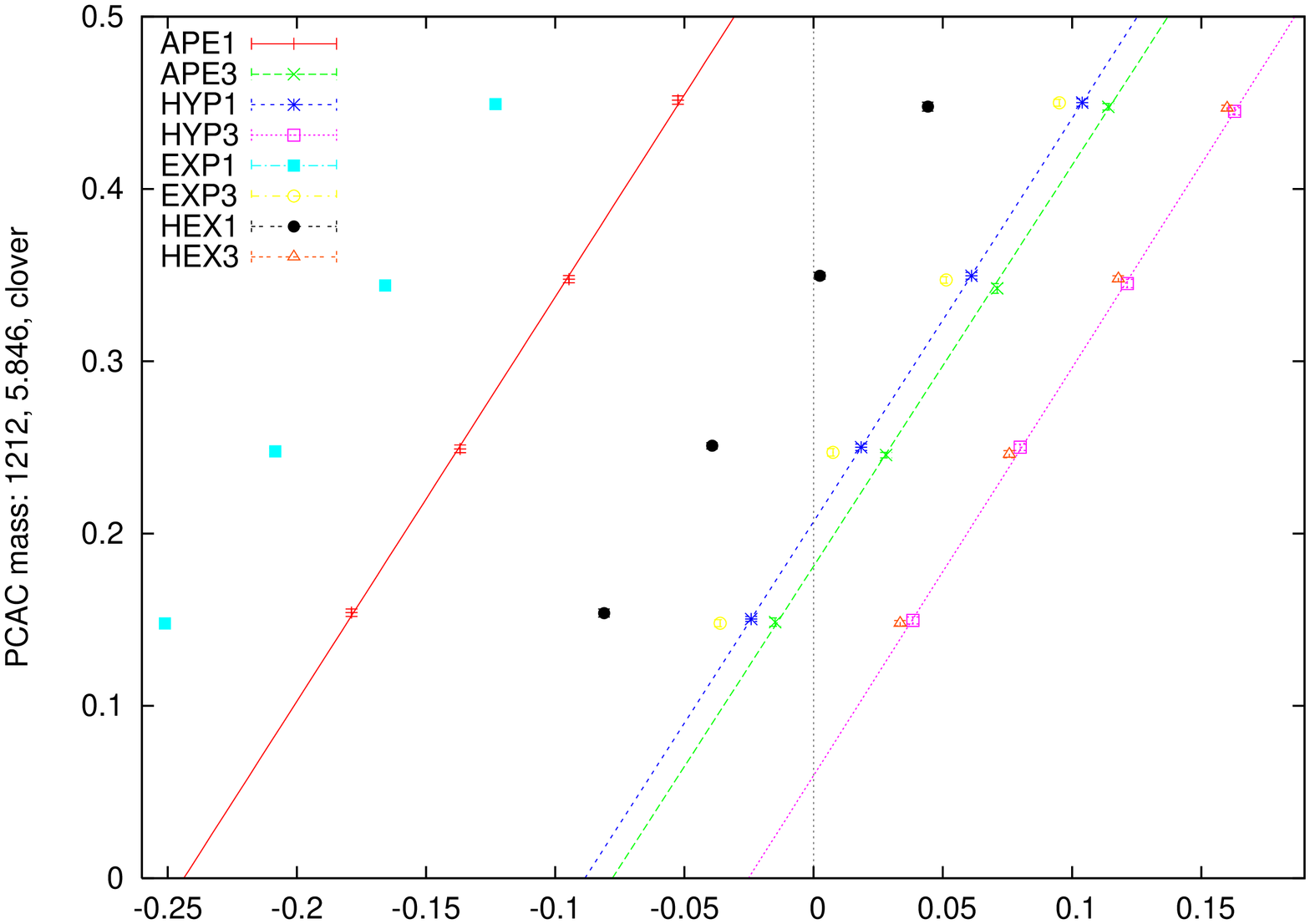,height=85mm,angle=-90}
\\[2mm]
\epsfig{file=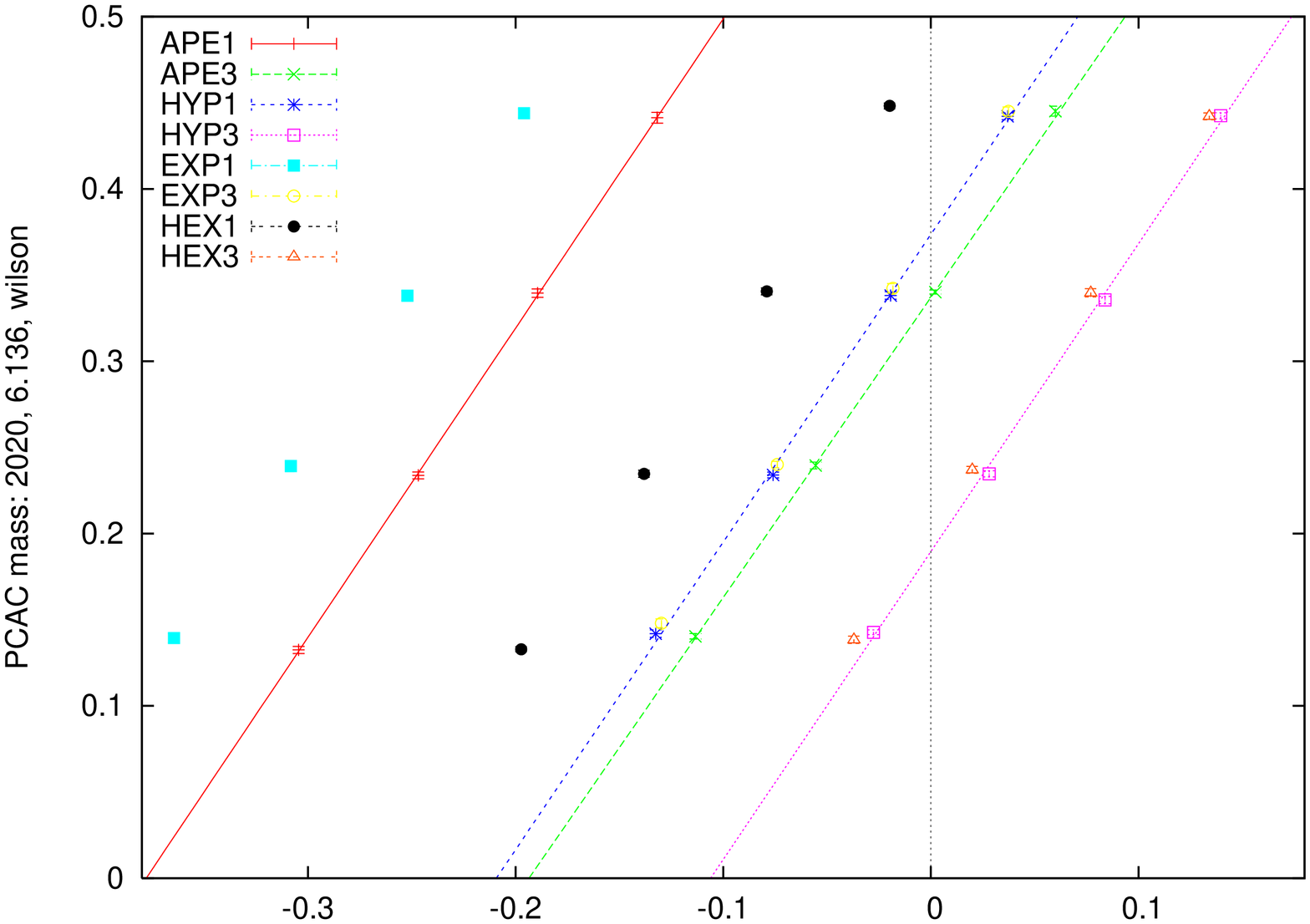,height=85mm,angle=-90}
\epsfig{file=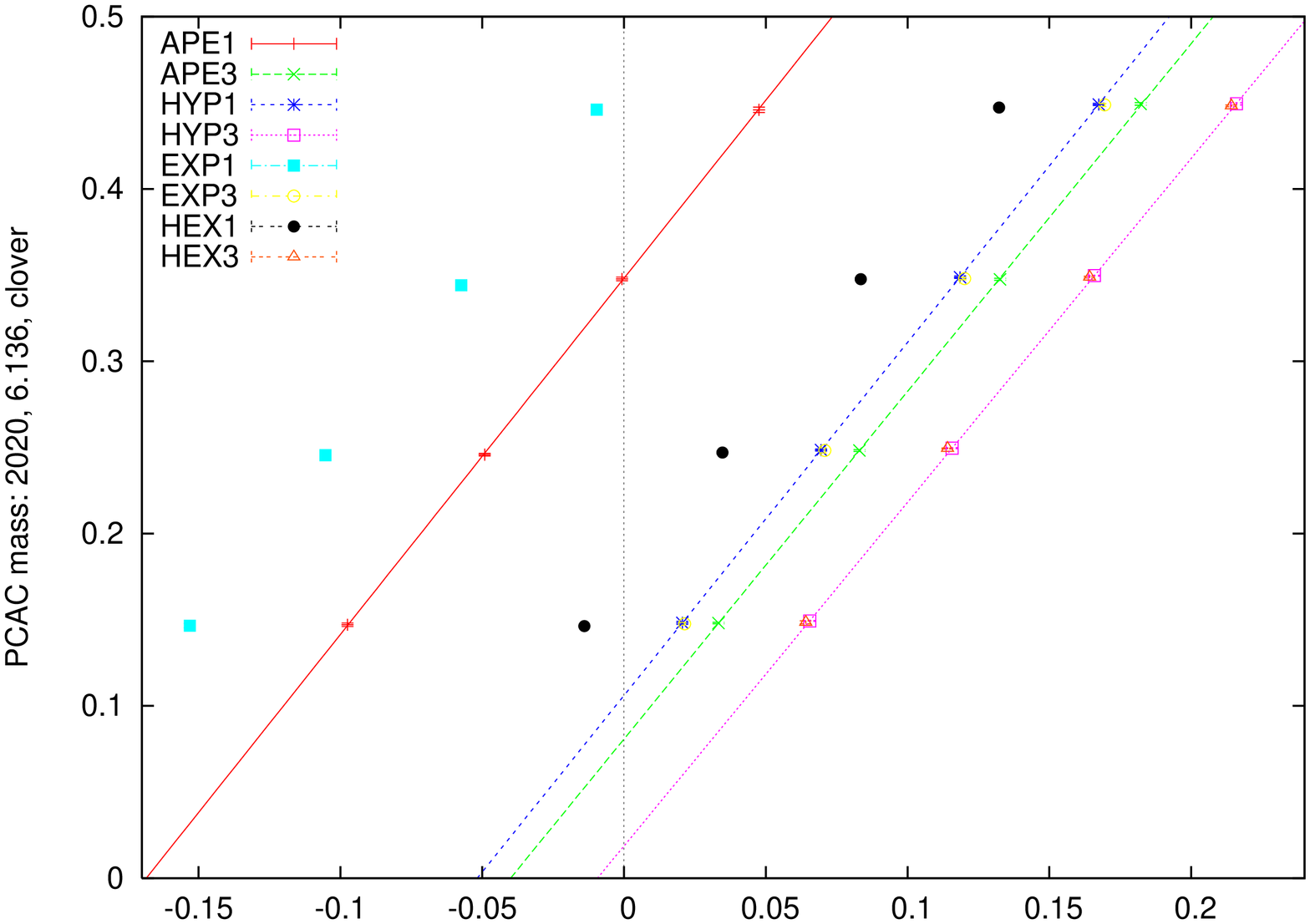,height=85mm,angle=-90}
\\[2mm]
\epsfig{file=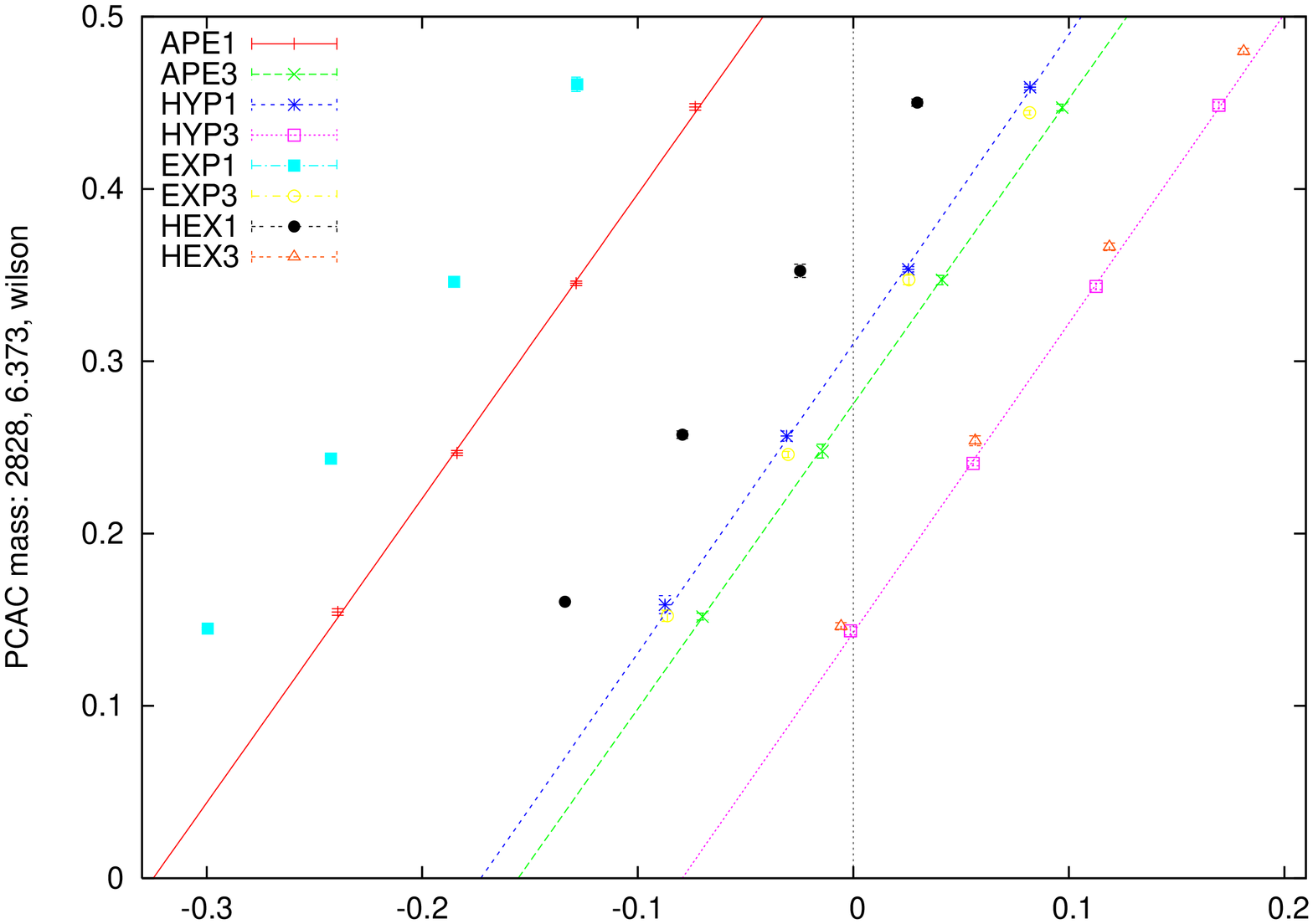,height=85mm,angle=-90}
\epsfig{file=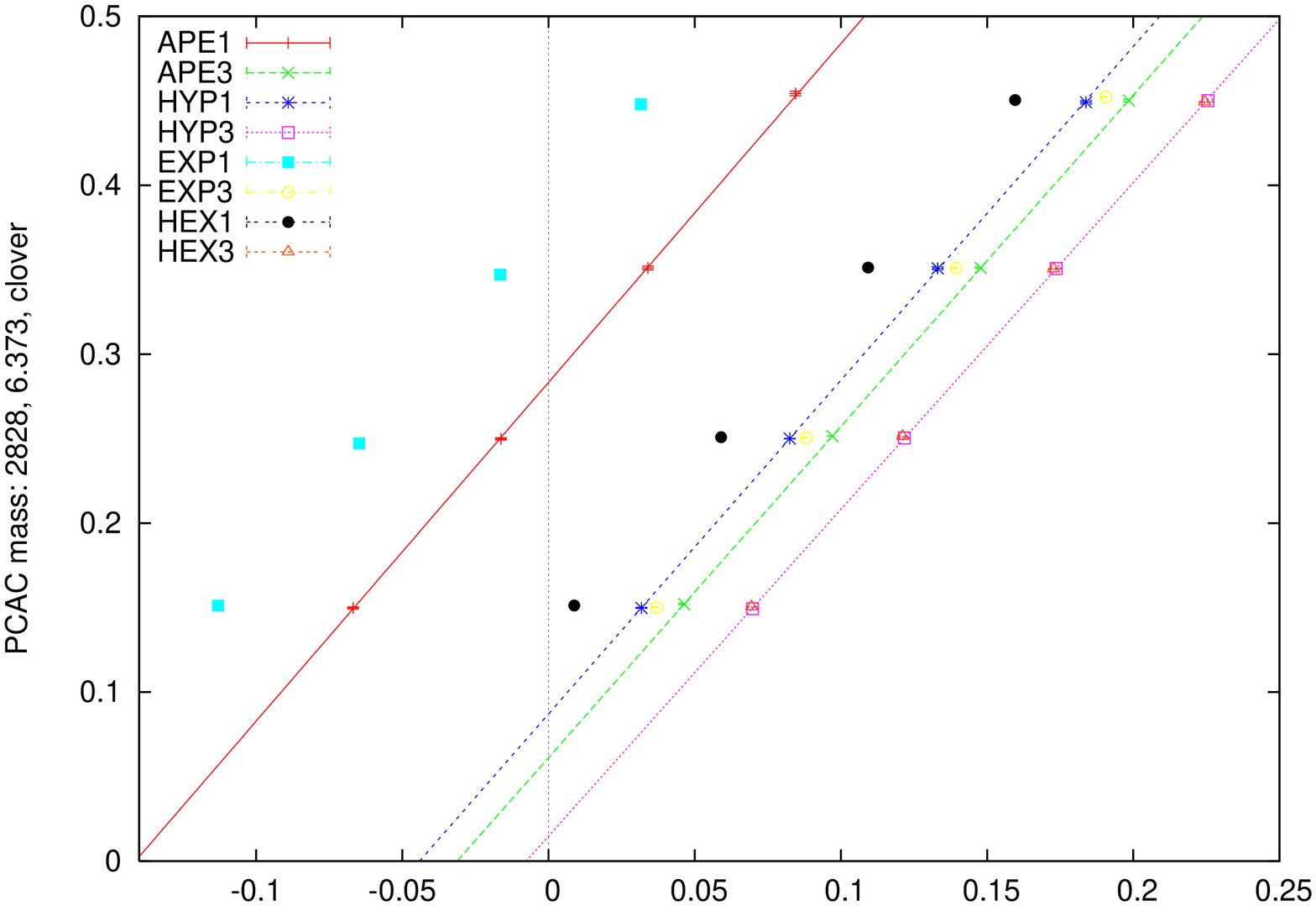,height=85mm,angle=-90}
\vspace{-2mm}
\caption{Our data for $2m^\mr{PCAC}$ vs.\ $m_0$ with $c_\mr{SW}\!=\!0$ (left)
and $c_\mr{SW}\!=\!1$ (right) at three couplings.}
\label{fig_mpcac_m0}
\end{figure}

For clover fermions one has, up to $O(ag_0^2,...,a^2)$ terms, the vector and
axial-vector Ward identities
\bea
Z_V\<.|\bar\pa_\mu \til V_\mu^a(x)|.\>\!&\!=\!&\!
{Z_m(a\mu)Z_S(a\mu)\ovr 4} (\til m_2^\mr{W}\!-\!\til m_1^\mr{W})
\<.|\til S^a(x\!+\!\hat4) +\!2\til S^a(x) +\!\til S^a(x\!-\!\hat4)|.\> \quad
\label{VWI}
\\
Z_A\<.|\bar\pa_\mu\til A_\mu^a(x)|.\>\!&\!=\!&\!
{Z_m(a\mu)Z_P(a\mu)\ovr 4} (\til m_2^\mr{W}\!+\!\til m_1^\mr{W})
\<.|\til P^a(x\!+\!\hat4)\!+\!2\til P^a(x)\!+\!\til P^a(x\!-\!\hat4)|.\> \quad
\label{AWI}
\eea
with $\til m^W\!=\!(1\!+\!b_mam^\mr{W})m^\mr{W}$.
The unmixed densities/currents $\til X$ with $X\!=\!S,P,V,A$ have been given
in (\ref{curr_impr}).
Note that either r.h.s.\ is scale-independent, since $Z_m\!=\!1/Z_S$ and the
two renormalization factors $Z_S$ and $Z_P$ run synchronously.
Finally, due to the $b_m$ term in (\ref{mquark_VWI}), $m_\mr{crit}$ does not
drop out of the r.h.s.\ of (\ref{VWI}) for unequal current quark masses.

A naive determination of $-am_\mr{crit}\!=\!4-1/(2\ka_\mr{crit})$ would measure
$M_\pi^2$ as a function of $m_0$ and determine, via an extrapolation, where the
former vanishes.
To avoid finite-volume and/or chiral log  effects, we determine $m^\mr{PCAC}$
as a function of $m_0$ and see where this quantity vanishes.
Up to $O(ag_0^2,...,a^2)$ [depending on the details of improvement] cut-off
effects this is a linear relationship, and, by virtue of (\ref{AWI}), the slope
is proportional to $Z_mZ_P/Z_A$.
More specifically, we restrict ourselves to degenerate quark masses (i.e.\
$m_1\!=\!m_2$) and employ the fitting ansatz
\beq
am^\mr{PCAC}={1\ovr\til Z_A}\,[1+b_m(am_0-am_\mr{crit})](am_0-am_\mr{crit})
\eeq
with $m_0$ the bare fermion mass given in (\ref{mquark_W}).
The goal is to test how well the fitted $-am_\mr{crit}$ and
$\til Z_A\!=\!Z_A/(Z_mZ_P)$ agree with the 1-loop prediction.
In principle, the coefficient $b_m$ is known at tree level.
It turns out that using this value leads to unacceptable fits.
On the other hand, our data are not precise enough to allow us to use $b_m$
as a parameter.
The quoted fits use $b_m\!=\!0$; this leads in most cases to acceptable
chisquares, and the few exceptions might be due to our limited statistics
(cf.\ Tab.\,\ref{tab_params}).
In fact, our data (taken at fixed $am^\mr{PCAC}$ to limit the CPU requirements)
do not show any visible curvature -- Fig.\,\ref{fig_mpcac_m0} shows the data
for three (out of five) couplings.
We performed several alternative fits (e.g.\ by dropping the last data point),
and as a result we estimate that the theoretical uncertainty is roughly one
order of magnitude larger than the statistical error quoted in
Tabs.\,\ref{tab_mcrit0}-\ref{tab_ZAtil1}.

\begin{table}
\begin{center}
\begin{tabular}{|c|c|c|c|c|c|}
\hline
$c_\mr{SW}\!=\!0$&$\be=5.846$&$\be=6.000$&$\be=6.136$&$\be=6.260$&$\be=6.373$\\
\hline
thin\,link       &  0.44573  &  0.43429  &  0.42466  &  0.41625  &  0.40887  \\
\hline
pert.\           &  0.11750  &  0.11448  &  0.11194  &  0.10973  &  0.10778  \\
1\,APE           & 0.5150(18)& 0.4283(17)& 0.3779(17)& 0.3496(22)& 0.3248(15)\\
1\,EXP           & 0.5846(23)& 0.4932(15)& 0.4412(21)& 0.4039(11)& 0.3805(09)\\
\hline
pert.\           &  0.04170  &  0.04063  &  0.03973  &  0.03894  &  0.03825  \\
3\,APE           & 0.2939(20)& 0.2247(16)& 0.1935(14)& 0.1685(16)& 0.1555(17)\\
3\,EXP           & 0.3263(21)& 0.2509(18)& 0.2100(19)& 0.1869(08)& 0.1713(19)\\
\hline
pert.\           &  0.06046  &  0.05891  &  0.05760  &  0.05646  &  0.05546  \\
1\,HYP           & 0.3094(18)& 0.2455(14)& 0.2093(16)& 0.1908(15)& 0.1728(28)\\
1\,HEX           & 0.3985(19)& 0.3158(18)& 0.2715(12)& 0.2449(16)& 0.2244(06)\\
\hline
pert.\           &    ---    &    ---    &    ---    &    ---    &    ---    \\
3\,HYP           & 0.1841(18)& 0.1290(14)& 0.1061(11)& 0.0949(17)& 0.0794(15)\\
3\,HEX           & 0.1993(18)& 0.1419(14)& 0.1142(16)& 0.0976(14)& 0.0868(16)\\
\hline
\end{tabular}
\caption{For $c_\mr{SW}\!=\!0$ Wilson fermions: $-am_\mr{crit}$ with 8
filtering recipes. In each field, the first row gives the (common) 1-loop
prediction, and the next two the linearly extrapolated values with APE/EXP or
HYP/HEX filtering, respectively. Errors are statistical only. We did not
measure the unfiltered $-am_\mr{crit}$, and we do not have a perturbative
prediction for 3\,HYP/HEX steps.}
\label{tab_mcrit0}
\end{center}
\begin{center}
\begin{tabular}{|c|c|c|c|c|c|}
\hline
$c_\mr{SW}\!=\!1$&$\be=5.846$&$\be=6.000$&$\be=6.136$&$\be=6.260$&$\be=6.373$\\
\hline
thin\,link       &  0.27719  &  0.27008  &  0.26409  &  0.25886  &  0.25427  \\
\hline
pert.\           &  0.04254  &  0.04145  &  0.04053  &  0.03973  &  0.03902  \\
1\,APE           & 0.2438(13)& 0.1929(08)& 0.1685(07)& 0.1518(08)& 0.1413(05)\\
1\,EXP           & 0.3140(13)& 0.2547(11)& 0.2231(08)& 0.2022(06)& 0.1873(04)\\
\hline
pert.\           &  0.00668  &  0.00651  &  0.00637  &  0.00624  &  0.00613  \\
3\,APE           & 0.0779(13)& 0.0497(07)& 0.0400(04)& 0.0341(02)& 0.0312(03)\\
3\,EXP           & 0.1003(13)& 0.0657(07)& 0.0512(06)& 0.0440(03)& 0.0392(02)\\
\hline
pert.\           &  0.01719  &  0.01675  &  0.01638  &  0.01605  &  0.01577  \\
1\,HYP           & 0.0885(10)& 0.0620(05)& 0.0517(05)& 0.0475(03)& 0.0441(03)\\
1\,HEX           & 0.1464(14)& 0.1045(07)& 0.0851(04)& 0.0743(03)& 0.0674(04)\\
\hline
pert.\           &    ---    &    ---    &    ---    &    ---    &    ---    \\
3\,HYP           & 0.0252(12)& 0.0120(05)& 0.0094(02)& 0.0084(01)& 0.0077(03)\\
3\,HEX           & 0.0289(10)& 0.0143(05)& 0.0111(04)& 0.0088(02)& 0.0088(02)\\
\hline
\end{tabular}
\caption{For $c_\mr{SW}\!=\!1$ clover fermions: $-am_\mr{crit}$ with 8
filtering recipes (cf.\ caption of Tab.\,\ref{tab_mcrit0}).}
\label{tab_mcrit1}
\end{center}
\end{table}

Our non-perturbative data for $-am_\mr{crit}$ are given in
Tab.\,\ref{tab_mcrit0} and Tab.\,\ref{tab_mcrit1} for the Wilson
($c_\mr{SW}\!=\!0$) and clover ($c_\mr{SW}\!=\!1$) case, respectively.
As an illustration, we add the 1-loop prediction that follows from
(\ref{def_amcrit}, \ref{mquark_W}) and Tab.\,\ref{tab_addshift}.
We did not measure the unfiltered $-am_\mr{crit}$, since it would be too
expensive for our computational resources, and the large discrepancy between
the perturbative and non-perturbative critical mass for unfiltered actions is
well known.

\begin{table}
\begin{center}
\begin{tabular}{|c|c|c|c|c|c|}
\hline
$c_\mr{SW}\!=\!0$&$\be=5.846$&$\be=6.000$&$\be=6.136$&$\be=6.260$&$\be=6.373$\\
\hline
thin\,link       &  0.94668  &  0.94805  &  0.94920  &  0.95020  &  0.95109  \\
\hline
pert.\           &  0.99859  &  0.99863  &  0.99866  &  0.99868  &  0.99871  \\
1\,APE           & 1.081(12) & 1.115(11) & 1.115(12) & 1.134(16) & 1.132(09) \\
1\,EXP           & 1.037(16) & 1.086(11) & 1.110(13) & 1.095(08) & 1.121(10) \\
\hline
pert.\           &  1.00281  &  1.00274  &  1.00268  &  1.00262  &  1.00258  \\
3\,APE           & 1.066(13) & 1.114(11) & 1.149(10) & 1.123(10) & 1.130(10) \\
3\,EXP           & 1.067(14) & 1.132(12) & 1.117(12) & 1.120(05) & 1.140(10) \\
\hline
pert.\           &  1.00061  &  1.00059  &  1.00058  &  1.00057  &  1.00056  \\
1\,HYP           & 1.046(12) & 1.129(09) & 1.120(10) & 1.137(09) & 1.114(14) \\
1\,HEX           & 1.070(12) & 1.117(12) & 1.126(08) & 1.139(09) & 1.129(07) \\
\hline
pert.\           &    ---    &    ---    &    ---    &    ---    &    ---    \\
3\,HYP           & 1.051(11) & 1.113(09) & 1.119(08) & 1.148(11) & 1.114(10) \\
3\,HEX           & 1.058(11) & 1.119(09) & 1.125(10) & 1.125(10) & 1.119(09) \\
\hline
\end{tabular}
\caption{For $c_\mr{SW}\!\!=\!0$ Wilson fermions: $Z_A/(Z_mZ_P)$ with 8
filtering recipes (cf.\ caption of Tab.\,\ref{tab_mcrit0}).}
\label{tab_ZAtil0}
\end{center}
\begin{center}
\begin{tabular}{|c|c|c|c|c|c|}
\hline
$c_\mr{SW}\!=\!1$&$\be=5.846$&$\be=6.000$&$\be=6.136$&$\be=6.260$&$\be=6.373$\\
\hline
thin\,link       &  0.90710  &  0.90949  &  0.91149  &  0.91324  &  0.91478  \\
\hline
pert.\           &  0.98030  &  0.98080  &  0.98123  &  0.98160  &  0.98193  \\
1\,APE           & 0.8523(86)& 0.9263(55)& 0.9679(51)& 0.9798(70)& 0.9974(42)\\
1\,EXP           & 0.8602(80)& 0.9477(44)& 0.9912(26)& 1.0079(17)& 1.0200(24)\\
\hline
pert.\           &  0.99424  &  0.99439  &  0.99451  &  0.99462  &  0.99472  \\
3\,APE           & 0.8554(69)& 0.9398(33)& 0.9761(31)& 1.0008(22)& 1.0126(24)\\
3\,EXP           & 0.8458(80)& 0.9503(28)& 1.0024(15)& 1.0258(10)& 1.0340(21)\\
\hline
pert.\           &  0.99014  &  0.99040  &  0.99061  &  0.99080  &  0.99096  \\
1\,HYP           & 0.8539(88)& 0.9195(72)& 0.9599(51)& 0.9699(49)& 0.9847(37)\\
1\,HEX           & 0.8703(78)& 0.9510(42)& 0.9841(34)& 1.0039(25)& 1.0160(15)\\
\hline
pert.\           &    ---    &    ---    &    ---    &    ---    &    ---    \\
3\,HYP           & 0.8517(93)& 0.9435(42)& 0.9713(33)& 0.9959(23)& 1.0074(28)\\
3\,HEX           & 0.8452(61)& 0.9548(30)& 1.0050(26)& 1.0193(13)& 1.0373(11)\\
\hline
\end{tabular}
\caption{For $c_\mr{SW}\!=\!1$ clover fermions: $Z_A/(Z_mZ_P)$ with 8
filtering recipes (cf.\ caption of Tab.\,\ref{tab_mcrit0}).}
\label{tab_ZAtil1}
\end{center}
\end{table}

Our non-perturbative data for $\til Z_A$ are given in Tab.\,\ref{tab_ZAtil0}
and Tab.\,\ref{tab_ZAtil1} for the cases $c_\mr{SW}\!=\!0$ and
$c_\mr{SW}\!=\!1$, respectively.
Note that $\til Z_A$ is scale-independent, since $Z_m\!=\!1/Z_S$, and the
factors $Z_S$ and $Z_P$ run synchronously.
Again, we add the 1-loop prediction that follows from (\ref{def_zX}) and
Tabs.\,\ref{tab_zX_wils}\,-\,\ref{tab_zX_ovim}.
For similar reasons as above, we did not measure the unfiltered $\til Z_A$.

The overall impression from Tabs.\,\ref{tab_mcrit0}-\ref{tab_ZAtil1} is that
1-loop perturbation theory does not give very accurate predictions for
non-perturbatively determined renormalization factors, if the improvement
coefficients are taken at tree-level.
However, the mismatch is much smaller if filtering \emph{and} improvement is
used -- as soon as one of the ingredients is missing, the ``agreement'' gets
much worse.
The virtue of the combined ``filtering and improvement'' program is that all
renormalization factors and improvement coefficients are close to their
respective tree-level values.
This is in marked contrast to other schemes (e.g.\ \cite{Lepage:1992xa}) in
which these quantities are far from $0$ and $1$, respectively, and the
challenge is to reproduce these big numbers in perturbation theory.


\subsection{Rational fits for $m_\mr{crit}$ with APE/HYP/EXP/HEX filtering}

\begin{figure}
\epsfig{file=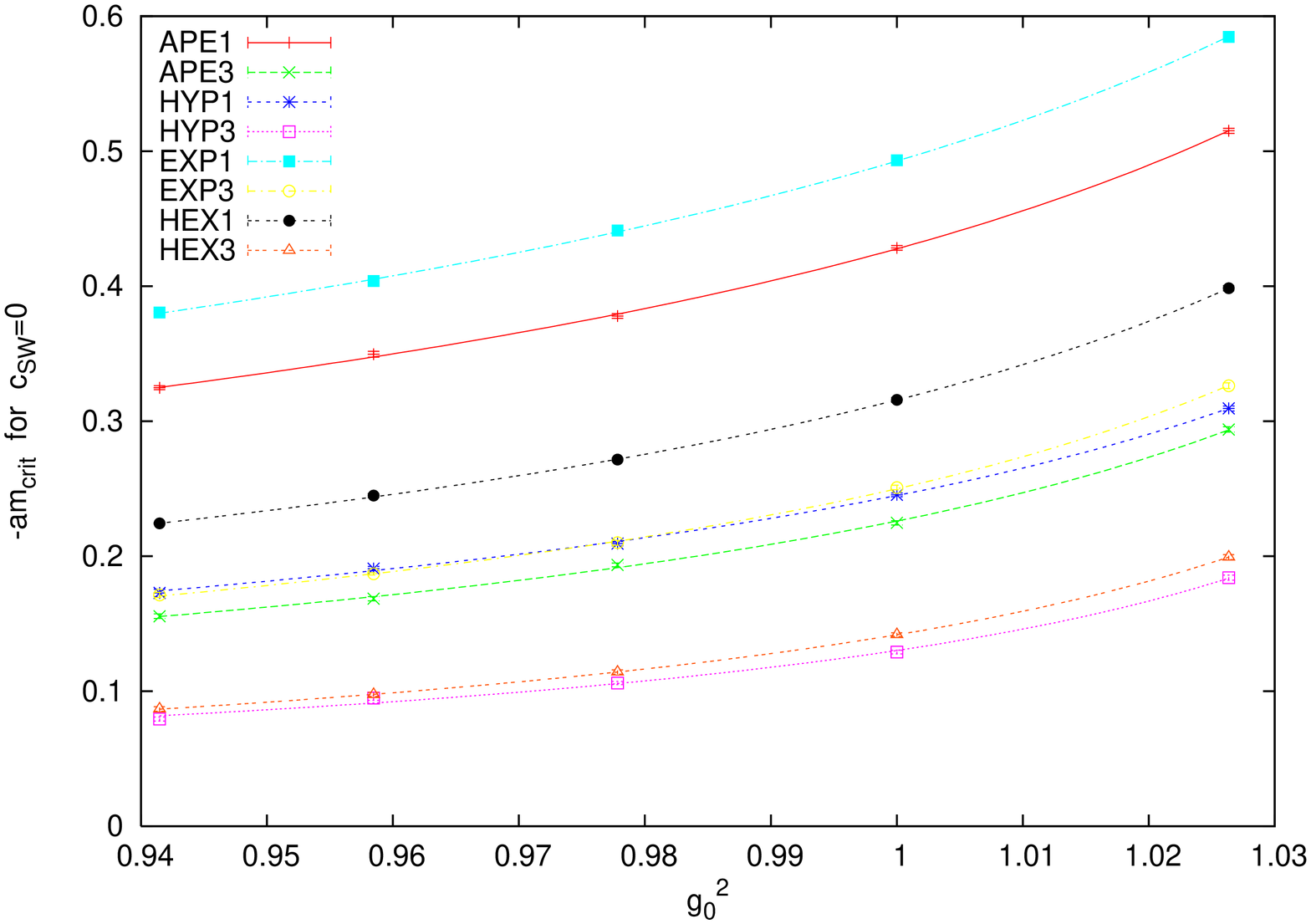,height=85mm,angle=-90}
\epsfig{file=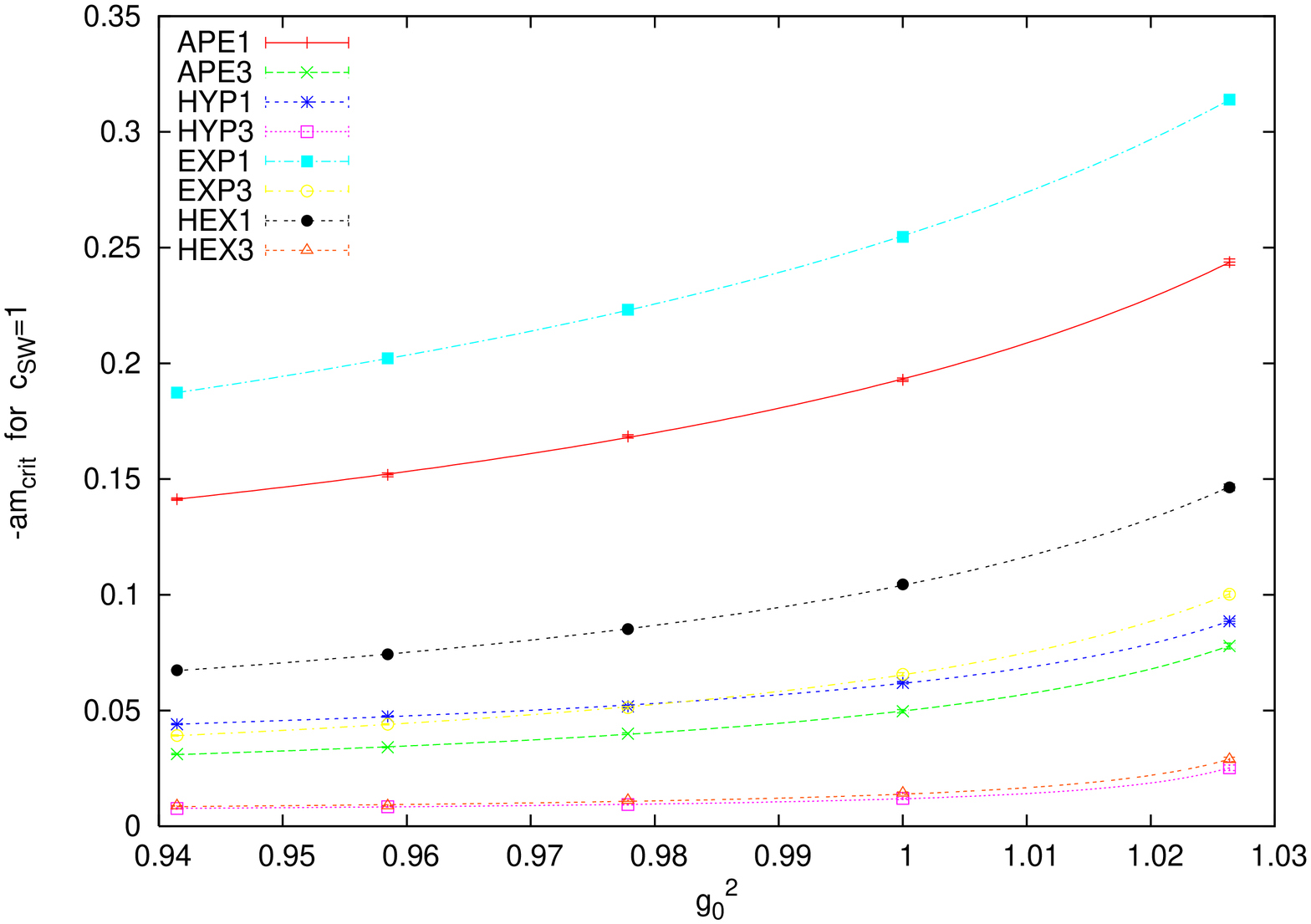,height=85mm,angle=-90}
\caption{$-am_\mr{crit}$ vs.\ $g_0^2$ for Wilson ($c_\mr{SW}\!=\!0$, left) and
clover ($c_\mr{SW}\!=\!1$, right) fermions with 8 filterings. The curves
indicate 3-parameter rational fits.}
\label{fig_rat_mcrit}
\end{figure}

\begin{table}
\begin{center}
\begin{tabular}{|cc|cc|cc|}
\hline
        &        & \twocolcc{$c_\mr{SW}=0$}& \twocolcc{$c_\mr{SW}=1$} \\
\hline
\twocolcc{pert.} &    \twocolcc{0.114480}  &    \twocolcc{0.0414467}  \\
 1\,APE & 1\,EXP &  0.213(12) & 0.252(12)  &  0.0909(28) & 0.1094(20) \\
\hline
\twocolcc{pert.} &    \twocolcc{0.040629}  &    \twocolcc{0.0065096}  \\
 3\,APE & 3\,EXP &  0.077(14) & 0.083(07)  &  0.0172(15) & 0.0171(09) \\
\hline
\twocolcc{pert.} &    \twocolcc{0.058906}  &    \twocolcc{0.0167502}  \\
 1\,HYP & 1\,HEX &  0.095(14) & 0.121(04)  &  0.0338(12) & 0.0332(16) \\
\hline
\twocolcc{pert.} &    \twocolcc{  ---   }  &    \twocolcc{  ---  }    \\
 3\,HYP & 3\,HEX &  0.034(15) & 0.026(01)  &  0.0060(02) & 0.0060(15) \\
\hline
\end{tabular}
\caption{The fitted coefficient $c_1$ in (\ref{rat_mcrit}), compared with the
1-loop prediction $S/(12\pi^2)$ with $S$ taken from Tab.\,\ref{tab_addshift}.}
\label{tab_rat_mcrit}
\end{center}
\end{table}

We know from (\ref{def_amcrit}) that asymptotically
$-am_\mr{crit} \to g_0^2S/(12\pi^2)=S/(2\pi^2\be)$ with $S$ given in
Tab.\,\ref{tab_addshift}.
Accordingly, if we fit our data with the rational ansatz
\beq
-am_\mr{crit}={c_1g_0^2+c_2g_0^4\ovr1+c_3g_0^2}
\label{rat_mcrit}
\eeq
then the coefficient $c_1$ would correspond, in the weak coupling regime,
to $S/(12\pi^2)$ with $S$ given in Tab.\,\ref{tab_addshift}.
Our data are not in the weak coupling regime, but still it is interesting to
check how much the coefficient $c_1$ from an unconstrained fit deviates from
the perturbative value.
The result is shown in Fig.\,\ref{fig_rat_mcrit} and Tab.\,\ref{tab_rat_mcrit}.
As was to be anticipated from our discussion of
Tabs.\ref{tab_mcrit0}-\ref{tab_mcrit1}, the ``agreement'' is not very good.
On an absolute scale the numbers are close, since they are all much smaller
than one.
On a relative scale, they deviate by a substantial factor.
In spite of this disagreement, the non-perturbative data still show a
consistency $c_1^\mr{APE}\!\simeq\!c_1^\mr{EXP}$ and ditto for
$c_1^\mr{HYP}\!\simeq\!c_1^\mr{HEX}$, as predicted in PT.
We find this amusing, in particular in view of the fact that the corresponding
(in PT) curves in Fig.\,\ref{fig_rat_mcrit} are not close at all.


\subsection{Rational fits for $\til Z_A$ with APE/HYP/EXP/HEX filtering}

\begin{figure}
\epsfig{file=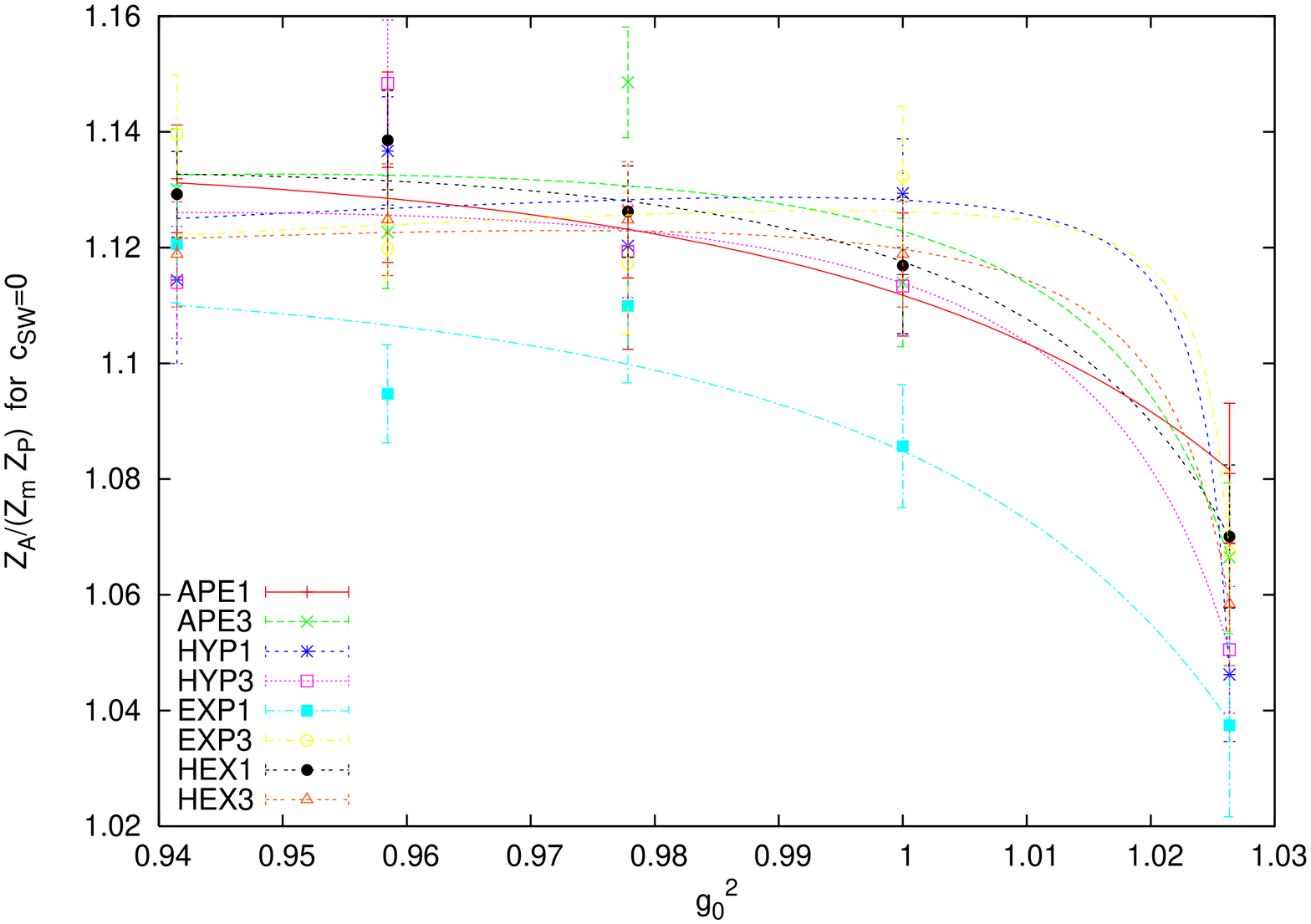,height=85mm,angle=-90}
\epsfig{file=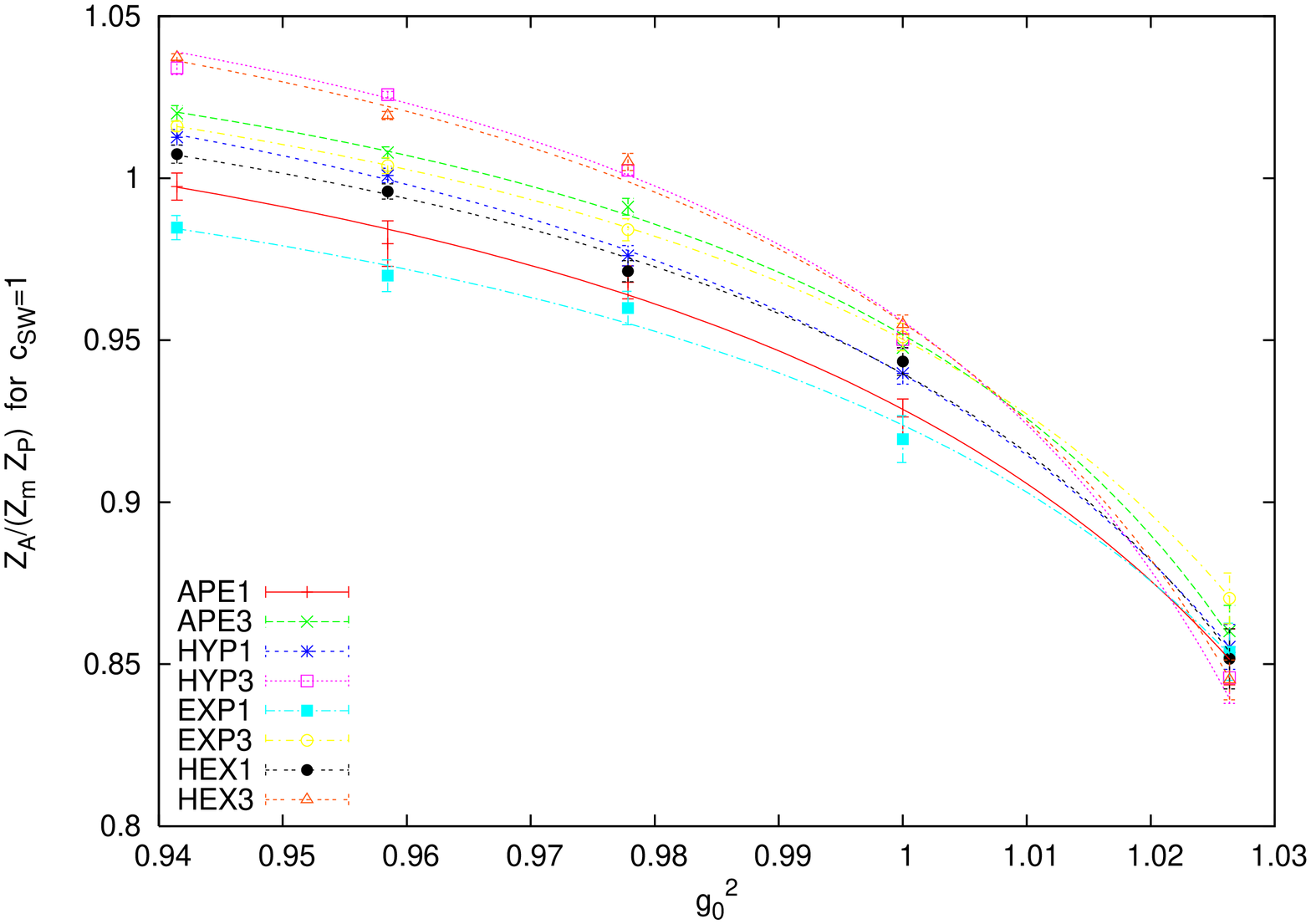,height=85mm,angle=-90}
\caption{$Z_A/(Z_mZ_P)$ vs.\ $g_0^2$ for Wilson ($c_\mr{SW}\!=\!0$, left) and
clover ($c_\mr{SW}\!=\!1$, right) fermions with 8 filterings. The curves
indicate 3-parameter rational fits.}
\label{fig_rat_zatil}
\end{figure}

We know from (\ref{def_zX}) that asymptotically
$\til Z_A \to 1\!-\!g_0^2(z_A\!+\!z_S\!-\!z_P)/(12\pi^2)=
1\!-\!(z_A\!+\!z_S\!-\!z_P)/(2\pi^2\be)$.
Accordingly, if we fit our data with the rational ansatz
\beq
\til Z_A={1+d_1g_0^2+d_2g_0^4\ovr1+d_3g_0^2}
\label{rat_zatil}
\eeq
then $d_1\!-\!d_3$ would correspond, in the weak coupling regime, to
$(z_A\!+\!z_S\!-\!z_P)/(12\pi^2)$ with $z_A,z_S,z_P$ given in
Tabs.\,\ref{tab_zX_wils}\,-\,\ref{tab_zX_ovim}.
The result of our fits is displayed in Fig.\,\ref{fig_rat_zatil}.
Again, there is no quantitative agreement between 1-loop perturbation theory
for $\til Z_A$ and our non-perturbative data, based on tree-level improvement
coefficients.
Still, comparing the two graphs in Fig.\,\ref{fig_rat_zatil}, one is led to
believe that with appropriate 1-loop improvement coefficients the situation
might be better.


\subsection{Rational fits for $m_\mr{res}$ with APE/HYP/EXP/HEX filtering}

We may express our result in terms of $m_\mr{res}\!=\!m^\mr{PCAC}(m_0\!=\!0)$.
$\til Z_A\!\simeq\!1$ implies $m_\mr{res}\!\simeq\!-m_\mr{crit}$, and we
refrain from copying Tabs.\,\ref{tab_mcrit0}-\ref{tab_mcrit1} with minimal
modifications.
Again, we performed rational fits, and the result looks very similar to
Fig.\,\ref{fig_rat_mcrit}.
An interesting observation is that $m_\mr{res}$ in physical units is almost
constant.
We find $m_\mr{res}^\mr{3\,APE}\!\simeq\!144,111,107,108,113\MeV$
at $\be\!=\!5.846,6.0,6.136,6.260,6.373$ and
$m_\mr{res}^\mr{3\,HYP}\!\simeq\!47,27,25,26,27\MeV$.
We feel confident that with 1-loop values for the coefficients
$c_\mr{SW},c_A,b_A\!-\!b_P$ smaller residual masses could be obtained.


\section{Summary}


We have presented a systematic study of filtered Wilson and clover quarks
in quenched QCD.
We have derived results at 1-loop order in weak-coupling perturbation theory
for $-am_\mr{crit}$ and the renormalization factors $Z_X$ with $X\!=\!S,P,V,A$
with four filterings [APE, HYP, EXP, HEX], in some cases with 1,2,3 iterations.
We have compared these predictions to non-perturbative data for $-am_\mr{crit}$
and $\til Z_A\!=\!Z_AZ_S/Z_P$ in a simulation without improvement and with
tree-level improvement coefficients.
We find no quantitative agreement in this specific setup.
Still, the tremendous progress that comes through the combination of tree-level
improvement and filtering leaves us optimistic that a theory with 1-loop
improvement coefficients and 2-loop renormalization factors might work in
practice.
By this we mean that a continuum extrapolation can be done from accessible
couplings as if the theory would have $O(a^2)$ cut-off effects only.

It turns out that lattice perturbation theory for UV-filtered fermion actions
is not much more complicated than for unfiltered actions.
For instance, our formula (\ref{app_b7}) gives a compact 1-loop expression for
the critical mass with an arbitrary number of APE smearings, and shows that
$am_\mr{crit}\to0$ for $n_\mr{iter}\to\infty$.
Since our results in the main part of the article were derived in a fully
automated manner, we feel that this explicit calculation provides an important
check.

One particularly compelling feature of filtered clover actions is that tadpole
resummation is not needed; in fact it barely changes the result.
This suggests that perturbation theory for filtered clover quarks converges
well.
In consequence, we expect that for filtered clover fermions the
non-perturbative improvement conditions as implemented by the ALPHA
collaboration \cite{ALPHA} will yield values consistent with such perturbative
predictions.

A beneficial feature in phenomenological applications is the low noise in
observables built from filtered clover quarks.
We have been able to determine $m_\mr{crit}$ to $\sim\!3\%$ statistical
accuracy from just a handful of configurations.
Therefore, the ``filtering'' comes at no cost -- it actually reduces the CPU
time needed to obtain a predefined accuracy in the continuum limit.

Let us comment on the filtering in two different fermion formulations.
It is clear that twisted-mass Wilson fermions would benefit from filtering,
too.
The dramatic renormalization of the twist angle would be tamed and it would be
much easier to realize maximum (renormalized) twist.
For rather different technical reasons, filtering has proven useful for overlap
fermions \cite{DeGrand:2002va,Kovacs:2002nz,Durr:2005an}.
In our technical study we decided to stay with $c_\mr{SW}\!=\!0$, because the
overlap prescription achieves automatic $O(a)$ on-shell improvement.
It is not clear to us whether the better chiral properties of a clover kernel
could translate into further savings in the overlap construction.

We hope that, once the 1-loop value for $c_\mr{SW}$ with $n$ iterations of the
EXP/stout recipe \cite{Morningstar:2003gk} is known%
\footnote{Note that at 2-loop order the strict correspondence between APE and
EXP with  $\al^\mr{APE}/6=\al^\mr{EXP}$ is lost.},
filtered clover fermions are ready for use in large-scale dynamical
simulations.
An important point is, of course, the smallest valence quark mass that can
be reached for a given coupling and sea quark mass (partially quenched setup).
We find $am_\mr{res}^\mr{3\,HYP}\!=\!0.0126(5)$ at $\be\!=\!6.0$
and $am_\mr{res}^\mr{3\,HYP}\!=\!0.0074(3)$ at $\be\!=\!6.373$ in the quenched
theory.
This corresponds to an almost constant residual mass in physical units,
$m_\mr{res}^\mr{3\,HYP}\!\simeq\!27\MeV$.
Since this mass is much smaller than in the unfiltered case, it is natural to
hope that one can reach smaller valence quark masses (in the quenched or
partially quenched setup) before one runs into the problem of ``exceptional''
configurations.
Furthermore, if mixing with unwanted chiralities in 4-fermi operators is an
$O((am_\mr{res})^2)$ effect \cite{Aoki:2005ga} in our case, too, the small
residual mass would be relevant for electroweak phenomenology.
Clearly, these topics deserve detailed investigations.

\subsubsection*{Acknowledgments}

We thank Tom DeGrand for useful correspondence.
S.D.\ is indebted to Ferenc Niedermayer for discussions on fat-link actions.
S.D.\ was supported by the Swiss NSF,
S.C.\ by the Fonds zur F\"orderung der Wissenschaftlichen Forschung in
\"Osterreich (FWF), Project P16310-N08.


\appendix\section{Fat link perturbation theory in $d$ dimensions}


\subsection{APE smearing}

In $d$ dimensions and with general gauge group $G$, standard APE smearing is
defined through
\bdm
U_\mu'(x)=P_G
\Big\{\,
(1\!-\!\al)U_\mu(x)+{\al\ovr2(d\!-\!1)}
\sum_{\pm\nu\neq(\mu)}
U_\nu(x)\,U_\mu(x\!+\!\hat\nu)\,U_\nu(x\!+\!\hat\mu)\dag
\Big\}
\edm
where the sum (``staple'') includes $2(d\!-\!1)$ terms.
The projection $P_G$ is needed, since in general the staple is no longer
a group element.
For the perturbative expansion we substitute
$U_\mu(x)\to1+\ri aA_\mu(x\!+\!{\hat\mu\ovr2})+O(a^2)$.
The prefactors $1\!-\!\al,\al/(2d\!-\!2)$ ensure that in PT the effect of $P_G$
is already taken care of.
For 2-quark and 4-quark renormalization factors at 1-loop order only the linear
part is relevant \cite{Bernard:1999kc}.
After shifting $x\to x\!-\!{\hat\mu\ovr2}$ one obtains at leading order
\bea
A_\mu'(x)&=&A_\mu(x)\;+\;
{\al\ovr2(d\!-\!1)}\sum_\nu
\Big\{A_\mu(x\!+\hat\nu)-2A_\mu(x)+A_\mu(x\!-\hat\nu)\Big\}
\nonumber
\\
&+&{\al\ovr2(d\!-\!1)}\sum_\nu
\Big\{
 A_\nu(x\!-\!{\hat\mu\ovr2}\!+\!{\hat\nu\ovr2})
-A_\nu(x\!-\!{\hat\mu\ovr2}\!-\!{\hat\nu\ovr2})
-A_\nu(x\!+\!{\hat\mu\ovr2}\!+\!{\hat\nu\ovr2})
+A_\nu(x\!+\!{\hat\mu\ovr2}\!-\!{\hat\nu\ovr2})
\Big\}
\quad
\label{ape_expanded}
\eea
where the sum now extends over all positive $\nu$.
This may be recast into the form
\bea
\om(y)&=&\de_{y,0}+
{\al\ovr2(d\!-\!1)}\sum_{\rh}\,
\{\de_{y,\hat\rh}-2\de_{y,0}+\de_{y,-\hat\rh}\}
\nonumber
\\
\om_{\mu\nu}(y)&=&{\al\ovr2(d\!-\!1)}
\Big[
 \de_{y,-{\hat\mu\ovr2}+{\hat\nu\ovr2}}-\de_{y,-{\hat\mu\ovr2}-{\hat\nu\ovr2}}
-\de_{y,+{\hat\mu\ovr2}+{\hat\nu\ovr2}}+\de_{y,+{\hat\mu\ovr2}-{\hat\nu\ovr2}}
\Big]
\nonumber
\\
A_\mu'(x)&=&\sum_{y,\nu}\,h_{\mu\nu}(y)\,A_\nu(x\!+\!y)=
\sum_{y,\nu}\,
\Big\{
[\om(y)\de_{\mu,\nu}+\om_{\mu\nu}(y)]\,A_\nu(x\!+\!y)
\Big\}
\eea
which is suitable for a Fourier transformation.
This leads to the final relation
\bea
A_\mu'(q)&=&
\sum\limits_\nu\,
\Big\{
\Big(
[1-{\al\ovr2(d\!-\!1)}\hat q^2]\de_{\mu,\nu}+
{\al\ovr2(d\!-\!1)}\hat q_\mu\hat q_\nu
\Big)\,A_\nu(q)
\Big\}
\nonumber
\\
&=&
[1-{\al\ovr2(d\!-\!1)}(\hat q^2\!-\!\hat q_\mu^2)]\,A_\mu(q)+
{\al\ovr2(d\!-\!1)}\sum\limits_{\nu\neq(\mu)}
\Big\{\hat q_\mu \hat q_\nu A_\nu(q)\Big\}
\nonumber
\\
&=&
A_\mu(q)+{\al\ovr2(d\!-\!1)}\sum\limits_{\nu\neq(\mu)}\,
\Big\{
-\hat q_\nu^2 A_\mu(q)+\hat q_\mu \hat q_\nu A_\nu(q)
\Big\}
\eea
with $\hat q_\rh={2\ovr a}\sin({a\ovr2}q_\rh)$ (for all $d$).
A form particularly useful for iterated smearing ($n\!>\!1$) is
\cite{Bernard:1999kc}
\beq
A_\mu^{(n)}(q)=\sum_\nu\,
\Big\{
\Big(
[1-{\al\ovr2(d\!-\!1)}\hat q^2]^n\,
(\de_{\mu,\nu}-{\hat q_\mu \hat q_\nu\ovr \hat q^2})
+{\hat q_\mu \hat q_\nu\ovr \hat q^2}
\Big)
\,A_\nu(q)
\Big\}
\label{ape_final}
\eeq
where the transverse part is simply a form-factor
$f^{(n)}(\qh^2)=[1-{\al\ovr2(d\!-\!1)}\hat q^2]^n$ as emphasized
in \cite{Bernard:1999kc}.


\subsection{HYP smearing}

In $d\!\geq\!3$ dimensions $d\!-\!1$ levels of restricted APE smearings may be
nested in such a way that the final ``fat'' link contains only ``thin'' links
in the adjacent hypercubes \cite{Hasenfratz:2001hp}.
Specifically, in $d\!=\!4$ the linearized HYP relation reads (note that
$\al_{3,2,1}$ refer to step 1,2,3, respectively)
\bea
\bar A_{\mu,\nu\rh}(x)&=&(1-\al_3)A_\mu(x)+
{\al_3\ovr2}\sum_{\pm\si\neq(\mu\nu\rh)}
\Big\{
A_\si(x\!-\!{\hat\mu\ovr2}\!+\!{\hat\si\ovr2})+
A_\mu(x\!+\!\si)+
A_\si(x\!+\!{\hat\mu\ovr2}\!+\!{\hat\si\ovr2})
\Big\}
\nonumber
\\
\til A_{\mu,\nu}(x)&=&(1-\al_2)A_\mu(x)+
{\al_2\ovr4}\sum_{\pm\rh\neq(\mu\nu)}
\Big\{
\bar A_{\rh,\mu\nu}(x\!-\!{\hat\mu\ovr2}\!+\!{\hat\rh\ovr2})+
\bar A_{\mu,\nu\rh}(x\!+\!\rh)+
\bar A_{\rh,\mu\nu}(x\!+\!{\hat\mu\ovr2}\!+\!{\hat\rh\ovr2})
\Big\}
\nonumber
\\
A_\mu'(x)&=&(1-\al_1)A_\mu(x)+
{\al_1\ovr6}\sum_{\pm\nu\neq(\mu)}
\Big\{
\til A_{\nu,\mu}(x\!-\!{\hat\mu\ovr2}\!+\!{\hat\nu\ovr2})+
\til A_{\mu,\nu}(x\!+\!\nu)+
\til A_{\nu,\mu}(x\!+\!{\hat\mu\ovr2}\!+\!{\hat\nu\ovr2})
\Big\}
\nonumber
\eea
and it is easy to see that the core recipe in each step is an APE smearing in
2,3,4 dimensions, respectively.
Therefore, the Fourier transform leads to the relations
\bea
\bar A_{\mu,\nu\rh}(q)&=&A_\mu(q)+
{\al_3\ovr2}\sum_{\si\neq(\mu\nu\rh)}
\Big\{
-\hat q_\si^2 A_\mu(q) +\hat q_\mu \hat q_\si A_\si(q)
\Big\}
\nonumber
\\
\til A_{\mu,\nu}(q)&=&(1-\al_2)A_\mu(q)+
{\al_2\ovr4}\sum_{\rh\neq(\mu\nu)}
\Big\{
(2-\hat q_\rh^2)\bar A_{\mu,\nu\rh}(q)+
\hat q_\mu \hat q_\rh \bar A_{\rh,\mu\nu}(q)
\Big\}
\nonumber
\\
A_\mu'(q)&=&(1-\al_1)A_\mu(q)+
{\al_1\ovr6}\sum_{\nu\neq(\mu)}
\Big\{
(2-\hat q_\nu^2)\til A_{\mu,\nu}(q)+
\hat q_\mu \hat q_\nu \til A_{\nu,\mu}(q)
\Big\}
\eea
where a simplification specific to the innermost level has been applied.
Plugging everything in we obtain a compact momentum space representation for
one level of HYP smearing
\bea
A_\mu'&=&
A_\mu+{\al_1\ovr6}\sum_{\nu\neq(\mu)}\Big\{
\qh_\mu\qh_\nu A_\nu-\qh_\nu^2A_\mu+
{\al_2\ovr4}\sum_{\rh\neq(\mu\nu)}\Big\{
2\qh_\mu\qh_\rh A_\rh-\qh_\rh^2[(2-\qh_\nu^2)A_\mu+\qh_\mu\qh_\nu A_\nu]+
\nonumber
\\
&{}&
{\al_3\ovr2}\sum_{\si\neq(\mu\nu\rh)}\Big\{
4\qh_\mu\qh_\si A_\si-\qh_\si^2[2\qh_\mu\qh_\rh A_\rh+
(2-\qh_\rh^2)[(2-\qh_\nu^2)A_\mu+\qh_\mu\qh_\nu A_\nu]]
\Big\}\Big\}\Big\}
\eea
which, however, entails some orthogonality constraints.
To get rid of the latter, we apply a number of tricks.
First, the sum over $\si$ is split into two parts.
The part quadratic in $\qh_\si$ can be made independent of the summation index
by virtue of $\qh_\si^2=\qh^2-\qh_\mu^2-\qh_\nu^2-\qh_\rh^2$.
Hence, what remains in the innermost summation is the term linear in $\qh_\si$.
This term, however, is independent of $\rh$, the next-level index.
Since the constraint lets it assume the other free value (after $\mu$ and $\nu$
have been fixed) than $\rh$, the total effect is the same as with
$\si\!\to\!\rh$ replaced, thus
\bea
A_\mu'&=&
A_\mu+{\al_1\ovr6}\sum_{\nu\neq(\mu)}\Big\{
\qh_\mu\qh_\nu A_\nu-\qh_\nu^2A_\mu+
{\al_2\ovr4}\sum_{\rh\neq(\mu\nu)}\Big\{
(2\!+\!\al_3(2-\qh^2+\qh_\mu^2+\qh_\nu^2+\qh_\rh^2))\qh_\mu\qh_\rh A_\rh
\nonumber
\\
&{}&
-[\qh_\rh^2+{\al_3\ovr2}(\qh^2-\qh_\mu^2-\qh_\nu^2-\qh_\rh^2)(2-\qh_\rh^2)]
[(2-\qh_\nu^2)A_\mu+\qh_\mu\qh_\nu A_\nu]
\Big\}\Big\}
\nonumber
\eea
is a representation with only two sums.
Next we pull out those parts which are independent of the index $\rh$.
Using $\sum_{\rh\neq(\mu\nu)}\qh_\rh^2=\qh^2-\qh_\mu^2-\qh_\nu^2$ in the
remainder yields
\bea
A_\mu'&=&
A_\mu+{\al_1\ovr6}\sum_{\nu\neq(\mu)}\Big\{
\qh_\mu\qh_\nu A_\nu-\qh_\nu^2A_\mu+
{\al_2\ovr4}\sum_{\rh\neq(\mu\nu)}\Big\{
(2\!+\!\al_3(2-\qh^2+\qh_\mu^2+\qh_\nu^2+\qh_\rh^2))\qh_\mu\qh_\rh A_\rh
\Big\}
\nonumber
\\
&{}&
-{\al_2\ovr4}
[(1\!+\!\al_3)(\qh^2-\qh_\mu^2-\qh_\nu^2)-
{\al_3\ovr2}[(\qh^2-\qh_\mu^2-\qh_\nu^2)^2-\sum_{\rh\neq(\mu\nu)}\qh_\rh^4]]
[(2-\qh_\nu^2)A_\mu+\qh_\mu\qh_\nu A_\nu]
\Big\}
\nonumber
\eea
where the bracket multiplying ${\al_3\ovr2}$ is just
$2\prod_{\rh\neq(\mu\nu)}\qh_\rh^2$.
Since a constrained product would be inconvenient for later use, we choose to
stay with the actual form, but now we relax the constraint on $\rh$ to differ
from $\mu$ only and compensate for the additional term.
This yields
\bea
A_\mu'&=&
A_\mu+{\al_1\ovr6}\sum_{\nu\neq(\mu)}\Big\{
[1-{\al_2\ovr4}(2+\al_3(2-\qh^2+\qh_\mu^2+2\qh_\nu^2))]\qh_\mu\qh_\nu A_\nu
-\qh_\nu^2A_\mu
\nonumber
\\
&{}&
-{\al_2\ovr4}
[(1+\al_3)(\qh^2-\qh_\mu^2-\qh_\nu^2)-{\al_3\ovr2}Q_{\mu\nu}]
[(2-\qh_\nu^2)A_\mu+\qh_\mu\qh_\nu A_\nu]
\nonumber
\\
&&
+{\al_2\al_3\ovr4}\qh_\nu^2\sum_{\rh\neq(\mu)}\Big\{\qh_\mu\qh_\rh A_\rh\Big\}
+{\al_2\ovr4}\sum_{\rh\neq(\mu)}\Big\{
(2\!+\!\al_3(2-\qh^2+\qh_\mu^2+\qh_\rh^2))\qh_\mu\qh_\rh A_\rh
\Big\}
\Big\}
\nonumber
\eea
with
$Q_{\mu\nu}=(\qh^2-\qh_\mu^2-\qh_\nu^2)^2-\sum_{\rh\neq(\mu\nu)}\qh_\rh^4$.
In the sum over $\rh$ the term which depends on $\nu$ has been isolated.
The reason is that the other term may be pulled out of the $\nu$-sum (this
yields a factor 3), and since the constraint is the same, renaming the index
$\rh\!\to\!\nu$ is then legal.
Applying a similar procedure to the $\nu$-independent factor of the former
term, we obtain the form
\bea
A_\mu'&=&
A_\mu+{\al_1\ovr6}\sum_{\nu\neq(\mu)}\Big\{
[1+\al_2(1+{\al_3\ovr4}(4-\qh^2+\qh_\mu^2+\qh_\nu^2))]\qh_\mu\qh_\nu A_\nu
-\qh_\nu^2A_\mu
\nonumber
\\
&{}&
-{\al_2\ovr4}
[(1+\al_3)(\qh^2-\qh_\mu^2-\qh_\nu^2)-{\al_3\ovr2}Q_{\mu\nu}]
[(2-\qh_\nu^2)A_\mu+\qh_\mu\qh_\nu A_\nu]
\Big\}
\nonumber
\eea
with just one sum [apart from the $A$-independent $\sum\qh_\rh^4$ in
$Q_{\mu\nu}=
(\qh^2-\qh_\mu^2-\qh_\nu^2)^2+\qh_\mu^4+\qh_\nu^4-\sum_\rh\qh_\rh^4$].
Now it takes a couple of algebraic manipulations to arrive at the form
\bea
A_\mu'&=&
A_\mu+{\al_1\ovr6}\sum_{\nu\neq(\mu)}\Big\{
[1+\al_2(1+\al_3-{1\ovr4}(1+2\al_3)(\qh^2-\qh_\mu^2-\qh_\nu^2)
+{\al_3\ovr8}Q_{\mu\nu})]\qh_\mu\qh_\nu A_\nu -\qh_\nu^2 A_\mu
\nonumber
\\
&{}&
-{\al_2\ovr4}
[(1+\al_3)(2\qh^2-2\qh_\mu^2-2\qh_\nu^2)
-\al_3((\qh^2)^2-2\qh^2\qh_\mu^2+2\qh_\mu^4-\sum_\rh\qh_\rh^4)
-(1-\al_3)\times
\nonumber
\\
&{}&
(\qh^2-\qh_\mu^2-\qh_\nu^2)\qh_\nu^2
+{\al_3\ovr2}((\qh^2)^2-2\qh^2\qh_\mu^2-2\qh^2\qh_\nu^2+
2\qh_\mu^4+2\qh_\mu^2\qh_\nu^2+2\qh_\nu^4-\sum_\rh\qh_\rh^4)\qh_\nu^2
]A_\mu
\Big\}
\nonumber
\eea
which is suitable to do the sum in the terms which are even in $q_\nu$.
This operation yields
\bea
A_\mu'&=&
A_\mu+{\al_1\ovr6}\sum_{\nu\neq(\mu)}\Big\{
[1+\al_2(1+\al_3-{1\ovr4}(1+2\al_3)(\qh^2-\qh_\mu^2-\qh_\nu^2)
+{\al_3\ovr8}Q_{\mu\nu})]\qh_\mu\qh_\nu A_\nu
\Big\}
\nonumber
\\
&{}&
-{\al_1\ovr6}(1+\al_2(1+\al_3))(\qh^2-\qh_\mu^2)A_\mu
+{\al_1\al_2\ovr24}
(1+2\al_3)((\qh^2)^2-2\qh^2\qh_\mu^2+2\qh_\mu^4-\sum_\rh\qh_\rh^4)A_\mu
\nonumber
\\
&{}&
-{\al_1\al_2\al_3\ovr48}
(((\qh^2)^2-2\qh^2\qh_\mu^2+4\qh_\mu^4-3\sum_\rh\qh_\rh^4)(\qh^2-\qh_\mu^2)
+2\sum_\rh\qh_\rh^6-2\qh_\mu^6)
A_\mu
\nonumber
\eea
and upon extending the sum and compensating for the additional term one finds
\bea
A_\mu'&=&
A_\mu+{\al_1\ovr6}\sum_\nu\Big\{
[1+\al_2(1+\al_3-{1\ovr4}(1+2\al_3)(\qh^2-\qh_\mu^2-\qh_\nu^2)
+{\al_3\ovr8}Q_{\mu\nu})]\qh_\mu\qh_\nu A_\nu
\Big\}
\nonumber
\\
&{}&
-{\al_1\ovr6}[1+\al_2(1+\al_3)]\qh^2A_\mu
+{\al_1\al_2\ovr24}
(1+2\al_3)[(\qh^2)^2-\qh^2\qh_\mu^2-\sum_\rh\qh_\rh^4]A_\mu
\nonumber
\\
&{}&
-{\al_1\al_2\al_3\ovr48}[
(\qh^2)^3-2(\qh^2)^2\qh_\mu^2+2\qh^2\qh_\mu^4
-3\qh^2\sum_\rh\qh_\rh^4+2\qh_\mu^2\sum_\rh\qh_\rh^4
+2\sum_\rh\qh_\rh^6
]A_\mu
\eea
which looks somewhat lengthy.
As was noted by DeGrand and collaborators
\cite{Hasenfratz:2001tw,DeGrand:2002va,DeGrand:2002vu}, defining
$\Omega_{\mu\nu}=1+\al_2(1+\al_3)-{\al_2\ovr4}((1+2\al_3)
(\qh^2-\qh_\mu^2-\qh_\nu^2)-{\al_3\ovr2}Q_{\mu\nu})$ allows for the compact
form
\beq
A_\mu'=
\sum_\nu\Big\{
\Big(1-{\al_1\ovr6}\sum_\rh\{\Omega_{\mu\rh}\qh_\rh^2\}\Big)\de_{\mu\nu}
+{\al_1\ovr6}\Omega_{\mu\nu}\qh_\mu\qh_\nu
\Big\}A_\nu
\eeq
without any constraint on $\nu$ or $\rh$.
The general form for iterated smearing ($n\!>\!1$) is
\beq
A_\mu^{(n)}=
\sum_\nu\Big\{
T_{\mu\nu}^{(n)}
\Big(\de_{\mu\nu}-{\qh_\mu\qh_\nu\ovr\qh^2}\Big)+
L_{\mu\nu}^{(n)}
{\qh_\mu\qh_\nu\ovr\qh^2}
\Big\}A_\nu
\label{hyp_final}
\eeq
with the transverse and the longitudinal form-factor both being the product
of $n$ factors with adjacent indices summed over and the first and last index
set to $\mu$ and $\nu$ respectively,
\bea
T_{\mu\nu}^{(n)}&=&\sum_{\la_1,...,\la_{n-1}}\prod_{i=1}^n
\Big(
1-{\al_1\ovr12}\sum_{\rh_i}
\{[\Omega_{\la_{i-1}\rh_i}\!+\!\Omega_{\la_i\rh_i}]\qh_{\rh_i}^2\}
\Big)\Big|_{\la_0=\mu,\la_n=\nu}
\\
L_{\mu\nu}^{(n)}&=&\sum_{\la_1,...,\la_{n-1}}\prod_{i=1}^n
\Big(
1-{\al_1\ovr12}\sum_{\rh_i}
\{[\Omega_{\la_{i-1}\rh_i}\!+\!\Omega_{\la_i\rh_i}]\qh_{\rh_i}^2\}
+{\al_1\ovr6}\Omega_{\la_{i-1}\la_i}\qh^2
\Big)\Big|_{\la_0=\mu,\la_n=\nu}
\;.
\eea
In practice only moderate $n$ are relevant, and for $n\!=\!2$ and $n\!=\!3$
the explicit formulae read
\bea
A_\mu^{(2)}&=&
\sum_\nu\Big\{
\Big(1-{\al_1\ovr6}\sum_\rh\{\Omega_{\mu\rh}\qh_\rh^2\}\Big)^2\de_{\mu\nu}+
\nonumber
\\
&&
\Big(
{\al_1\ovr6}\Omega_{\mu\nu}(2-
{\al_1\ovr6}\sum_\rh
\{[\Omega_{\mu\rh}\!+\!\Omega_{\nu\rh}]\qh_\rh^2\})+
{\al_1^2\ovr36}\sum_\rh
\{\Omega_{\mu\rh}\Omega_{\rh\nu}\qh_\rh^2\}
\Big)\qh_\mu\qh_\nu
\Big\}A_\nu
\label{hyp_nis2}
\\[2mm]
A_\mu^{(3)}&=&
\sum_\nu\Big\{
\Big(1-{\al_1\ovr6}\sum_\rh\{\Omega_{\mu\rh}\qh_\rh^2\}\Big)^3\de_{\mu\nu}+
\Big(
{\al_1\ovr6}\Omega_{\mu\nu}
\Big[
3-{\al_1\ovr2}\sum_\rh\{[\Omega_{\mu\rh}\!+\!\Omega_{\nu\rh}]\qh_\rh^2\}
\nonumber
\\
&&
+{\al_1^2\ovr36}(\sum_\rh\{[\Omega_{\mu\rh}\!+\!\Omega_{\nu\rh}]\qh_\rh^2\})^2
-{\al_1^2\ovr36}
\sum_\rh\{\Omega_{\mu\rh}\qh_\rh^2\}\sum_\la\{\Omega_{\nu\la}\qh_\la^2\}
\Big]
\nonumber
\\
&&
+{\al_1^2\ovr36}\sum_\rh\{\Omega_{\mu\rh}\Omega_{\rh\nu}
(3-{\al_1\ovr6}\sum_\la
\{[\Omega_{\mu\la}\!+\!\Omega_{\nu\la}\!+\!\Omega_{\rh\la}]
\qh_\la^2\})\qh_\rh^2\}
\nonumber
\\
&&
+{\al_1^3\ovr216}\sum_{\rh,\la}
\{
\Omega_{\mu\rh}\Omega_{\rh\la}\Omega_{\la\nu}\qh_\rh^2\qh_\la^2
\}
\Big)\qh_\mu\qh_\nu
\Big\}A_\nu
\label{hyp_nis3}
\eea
but it is still clear that in general the transverse part contains a factor
$(1-{\al_1\ovr6}\sum_\rh\{\Omega_{\mu\rh}\qh_\rh^2\})^n$.


\subsection{EXP smearing}

Here we consider the EXP/stout smearing $U_\mu'(x)=S_\mu(x)U_\mu(x)$ [no sum]
introduced in \cite{Morningstar:2003gk} with
\bdm
S_\mu(x)\!=\!\exp\Big(
{\al\ovr2}\Big\{
\Big[\!\sum_{\pm\nu\neq(\mu)}\!
U_\nu(x)U_\mu(x\!+\!\hat\nu)U_\nu\dag(x\!+\!\hat\mu)U_\mu\dag(x)-
U_\mu(x)U_\nu(x\!+\!\hat\mu)U_\mu\dag(x\!+\!\hat\nu)U_\nu\dag(x)
\Big]-{1\ovr3}\mr{Tr}[.]
\Big\}
\Big)
\edm
a special unitary matrix by construction.
Upon expanding as before we obtain
\bdm
1+\ri aA_\mu'(x)=\Big(
1+\ri a\al\!\sum_{\pm\nu\neq(\mu)}\!
\Big\{
A_\nu(x\!-\!{\hat\mu\ovr2}\!+\!{\hat\nu\ovr2})+
A_\mu(x\!+\!\hat\nu)-
A_\nu(x\!+\!{\hat\mu\ovr2}\!+\!{\hat\nu\ovr2})-
A_\mu(x)
\Big\}
\Big)\Big(1+\ri aA_\mu(x)\Big)
\edm
and thus (still, up to terms of order $O(a^2)$)
\bea
A_\mu'(x)&=&
(1-2(d\!-\!1)\al)A_\mu(x)+
\al\sum_{\pm\nu\neq(\mu)}
\Big\{
A_\nu(x\!-\!{\hat\mu\ovr2}\!+\!{\hat\nu\ovr2})+
A_\mu(x\!+\!\hat\nu)-
A_\nu(x\!+\!{\hat\mu\ovr2}\!+\!{\hat\nu\ovr2})
\Big\}
\nonumber
\\
&=&A_\mu(x)\;+\;
\al\sum_\nu
\Big\{A_\mu(x\!+\hat\nu)-2A_\mu(x)+A_\mu(x\!-\hat\nu)\Big\}
\nonumber
\\
&+&\al\sum_\nu
\Big\{
 A_\nu(x\!-\!{\hat\mu\ovr2}\!+\!{\hat\nu\ovr2})
-A_\nu(x\!-\!{\hat\mu\ovr2}\!-\!{\hat\nu\ovr2})
-A_\nu(x\!+\!{\hat\mu\ovr2}\!+\!{\hat\nu\ovr2})
+A_\nu(x\!+\!{\hat\mu\ovr2}\!-\!{\hat\nu\ovr2})
\Big\}
\quad
\eea
which is just (\ref{ape_expanded}) with a modified parameter.
Accordingly, 1-loop fat link perturbation theory for EXP/stout smearing follows
from the version for APE smearing through the replacement
\beq
\al^\mr{APE}\;\longrightarrow\;2(d\!-\!1)\al^\mr{EXP/stout}
\;.
\label{replace_exp}
\eeq


\subsection{HEX smearing}

A natural generalization of the HYP concept is to use EXP/stout smearing in
each of the 3 steps (in 4D) rather than the standard APE smearing
\cite{Hasenfratz:2001hp}.
This entails the general definition
\bea
\bar V_{\mu,\nu\rh} (x)\!&\!=\!&\!\exp
\Big({\al_3\ovr2}\Big\{\Big[
\!\!\!\!\sum_{\pm\si\neq(\mu,\nu,\rh)}\!\!\!\!
U^{(n-1)}_{\si}(x)
U^{(n-1)}_{\mu}(x\!+\!\hat\si)
U^{(n-1)}_{\si}(x\!+\!\hat\mu)\dag
U^{(n-1)}_{\mu}(x)\dag
-\mr{h.c.}\Big]-{1\ovr3}\mr{Tr}[.]\Big\}
\Big)
\nonumber\\
&&
U_\mu^{(n-1)}(x)
\nonumber\\
\til V_{\mu,\nu} (x)\!&\!=\!&\!\exp
\Big({\al_2\ovr2}\Big\{\Big[
\!\!\sum_{\pm\si\neq(\mu,\nu)}\!\!
\bar V_{\si,\mu\nu}(x)
\bar V_{\mu,\nu\si}(x\!+\!\hat\si)
\bar V_{\si,\mu\nu}(x\!+\!\hat\mu)\dag
U^{(n-1)}_{\mu}(x)\dag 
-\mr{h.c.}\Big]-{1\ovr3}\mr{Tr}[.]\Big\}
\Big)
\nonumber\\
&&U_\mu^{(n-1)}(x)
\nonumber\\
U_\mu^{(n)}(x)\!&\!=\!&\!\exp
\Big({\al_1\ovr2}\Big\{\Big[
\sum_{\pm\nu\neq(\mu)}
\til V_{\nu,\mu}(x)
\til V_{\mu,\nu}(x\!+\!\hat\nu)
\til V_{\nu,\mu}(x\!+\!\hat\mu)\dag
U^{(n-1)}_{\mu}(x)\dag 
-\mr{h.c.}\Big]-{1\ovr3}\mr{Tr}[.]\Big\}
\Big)
\nonumber\\
&&
U_\mu^{(n-1)}(x)
\label{def_hex}
\eea
where again $\al_{3,2,1}$ refer to step 1,2,3, respectively, and no summation
over $\mu$ is implied.
We refer to (\ref{def_hex}) as ``hypercubically nested EXP'' or ``HEX''
smearing.
With (\ref{replace_exp}) it follows that
\beq
( \al_1^\mr{HYP}, \al_2^\mr{HYP}, \al_3^\mr{HYP})
\;\longrightarrow\;
(6\al_1^\mr{HEX},4\al_2^\mr{HEX},2\al_3^\mr{HEX})
\label{replace_hex}
\eeq
will automatically generate the perturbative formulae for the HEX recipe
(\ref{def_hex}).


\subsection{Permissible parameter ranges}

Regarding a reasonable range of smearing parameters, a standard criterion that
one may impose to avoid instabilities at higher iteration levels
is that the form-factor shall be smaller than 1 in absolute magnitude over the
entire Brillouin zone.
Since $\qh^2\!\leq\!4d$, formula (\ref{ape_final}) gives
\beq
\al_\mr{max}^\mr{APE}={d-1\ovr d}
\label{cond_APE}
\eeq
for APE smearing with arbitrary iteration number $n$.
With the replacement prescription (\ref{replace_exp}) the analogous condition
for EXP/stout smearing is $\al^\mr{EXP}\!\leq\!{1\ovr2d}$.

For $n$ HYP smearings in 4D the transverse part contains the
factor $(1-{\al_1\ovr6}\sum_\rh\{\Omega_{\mu\rh}\qh_\rh^2\})^n$, and requiring
this to be bounded in absolute magnitude by 1 leads to the two-fold condition
\bdm
0\leq
\sum_\rh\{
\al_1(1+\al_2(1+\al_3)-{\al_2\ovr4}[(1+2\al_3)
(\qh^2-\qh_\mu^2-\qh_\rh^2)-{\al_3\ovr2}
[(\qh^2-\qh_\mu^2-\qh_\rh^2)^2+\qh_\mu^4+\qh_\rh^4-\sum_\la\qh_\la^4]])
\qh_\rh^2\}\leq12
\edm
for each $\mu$.
Accordingly, upon summing everything over $\mu$ one finds
\bdm
0\leq\sum_\rh\{
\al_1(4+4\al_2(1+\al_3)-{\al_2\ovr4}[(1+2\al_3)
(3\qh^2-4\qh_\rh^2)-{\al_3\ovr2}
[2(\qh^2)^2-6\qh^2\qh_\rh^2+8\qh_\rh^4
-2\sum_\la\qh_\la^4]])
\qh_\rh^2\}
\leq48
\edm
and then doing the sum over $\rh$ yields the inequality
\bdm
0\leq
4\al_1(1+\al_2(1+\al_3))\qh^2
-{\al_1\al_2\ovr4}[(1+2\al_3)
(3(\qh^2)^2-4\sum_\la\qh_\la^4)-\al_3
[(\qh^2)^3-4\qh^2\sum_\la\qh_\la^4+4\sum_\la\qh_\la^6]]
\leq48
\edm
which is a non-trivial constraint on
$(\al_1^\mr{HYP},\al_2^\mr{HYP},\al_3^\mr{HYP})$ in terms of the three
quantities
\bdm
0\leq\sum_\la\qh_\la^2\leq16
\quad,\qquad
0\leq\sum_\la\qh_\la^4\leq64
\quad,\qquad
0\leq\sum_\la\qh_\la^6\leq256
\edm
but the latter are, of course, not independent.
Neglecting this, the condition
\beq
0\leq
\al_1(1+\al_2(1+\al_3))[0...64]
+\al_1\al_2(1+2\al_3)[-192...64]
+\al_1\al_2\al_3[-1024...1280]
\leq48
\eeq
can be separated into one for the lower and one for the upper bound.
While the former is always satisfied for positive smearing parameters, the
latter takes the form
\beq
\al_1^\mr{HYP}(1+\al_2^\mr{HYP}(2+23\al_3^\mr{HYP}))
\leq{3\ovr4}
\;.
\label{HYP_cond1}
\eeq
Another useful form might arise from keeping only the part quadratic in
the momenta in the inequality, as the remainder may have either sign, and this
leads to the less restrictive condition
\beq
\al_1^\mr{HYP}(1+\al_2^\mr{HYP}(1+\al_3^\mr{HYP}))
\leq{3\ovr4}
\;.
\label{HYP_cond2}
\eeq
Note that for $\al_2\!=\al_3\!=\!0$ either condition coincides with
(\ref{cond_APE}).
Finally, we mention that neither (\ref{HYP_cond1}) nor (\ref{HYP_cond2}) is
satisfied by the standard HYP parameter set (\ref{std_HYP_HEX}).
Note, however, that these are not necessary conditions; they emerged from
applying some simplifications to a highly non-linear precessor.
Applying the replacement recipe (\ref{replace_hex}), the
analogous conditions for HEX smearing are found to be
$\al_1^\mr{HEX}(1+8\al_2^\mr{HEX}(1+23\al_3^\mr{HEX}))\!\leq\!{1\ovr8}$
and
$\al_1^\mr{HEX}(1+4\al_2^\mr{HEX}(1+2\al_3^\mr{HEX}))\!\leq\!{1\ovr8}$,
respectively.


\subsection{Diffusion law for iterated smearing}

As a consequence of (\ref{ape_final}), the form-factor for the transverse part
after $n$ APE smearings is \cite{Bernard:1999kc}
\beq
f^{(n)}(\qh^2)\simeq\exp(-{n\,\al^\mr{APE}\ovr2(d-1)}\qh^2)+O((\qh^2)^2)
\;.
\eeq
This means that the square-radius of the resulting form-factor takes the form
\beq
\<r^2\>_\mr{APE}={n\,\al^\mr{APE}\ovr d-1}
\label{footprint_ape}
\eeq
which is a diffusion law, since the smearing effectively affects a space-time
region growing like $\<r^2\>_\mr{APE}^{1/2}\!\propto\!\sqrt{n}$.
Focusing on the quadratic part in the transverse factor below (\ref{hyp_nis3})
one finds
\beq
\<r^2\>_\mr{HYP}={n\,\al_1\ovr3}(1+\al_2(1+\al_3))
\label{footprint_hyp}
\eeq
for $n$ iterations of HYP smearing in 4D.
As noted in \cite{Bernard:1999kc}, the prefactors are favorably small.
Even 3 APE steps with $\al_\mr{std}^\mr{APE}$ generate a ``footprint''
$\<r^2\>_\mr{APE}^{1/2}\!\simeq\!0.775$ i.e.\ of the order of one lattice
spacing.
Likewise, 3 HYP smearings with $\al_\mr{std}^\mr{HYP}$ yield
$\<r^2\>_\mr{HYP}^{1/2}\!\simeq\!1.155$.


\section{Additive mass renormalization with filtering}


Here we give a derivation of the additive mass renormalization for APE-filtered
clover fermions at 1-loop order in lattice perturbation theory.
We work in Feynman gauge; the effect of smearing is not just a modification of
the gluon propagator, as it is in Landau gauge \cite{Bernard:1999kc}.

For the gauge field we use the same conventions as in App.\,A, that is
\bea
A_\mu^{(n)}(q)&=&\til h_{\mu\nu}^{(n)}(q) \, A_\mu^{(0)}(q)
\\
\til h_{\mu\nu}^{(n)}(q)&=&
(1-{\al\ovr6})^n\,(\de_{\mu\nu}-{\qh_\mu\qh_\nu\ovr\qh^2})+
{\qh_\mu\qh_\nu\ovr\qh^2}
=f^n(q)\de_{\mu\nu}-(f^n(q)\!-\!1){\qh_\mu\qh_\nu\ovr\qh^2}
\eea
with $f(q)\!=\!1\!-\!(\al/6)\qh^2$ and $\qh\!=\!2\sin(q_\mu/2)$, except that
repeated indices are always summed over in this appendix.
Furthermore, we use the shorthand notation
\bea
s_\mu=\sin({q_\mu\ovr2})\;, &  & s^2=s_\mu s_\mu
\nonumber\\
\bar{s}_\mu=\sin(q_\mu)\;,  &  & \bar{s}^2=\bar{s}_\mu \bar{s}_\mu
\nonumber
\eea
and analogously $c_\mu\!=\!\cos(q_\mu/2)$ and $c^2\!=\!c_\mu c_\mu$ with
summation implicit.

With these conventions the gluon and quark propagators (in Feynman gauge) take
the form
\bea
G_{\mu\nu}(q)&=&\de_{\mu\nu}\,G(q) \;,\qquad G(q)={1\ovr4s^2}
\\
S(q)&=&{B(q)\ovr\Delta(q)}=
{2s^2-\ri\ga_\mu\bar{s}_\mu\ovr4(s^2)^2+\bar{s}^2}
\eea
and the two-quark (zero external momentum on one side) one-gluon coupling is
$V_\rh\!\pm\!W_\rh$ with
\bea
V_\rh(q)&=&-\ri\ga_\rh c_\rh-s_\rh
\label{app_b1}
\\
W_\rh(q)&=&-{c_\mr{SW}\ovr2\ri}\si_{\rh\la}c_\rh\bar{s}_\la
\qquad\mbox{(sum over $\la$ only)}
\label{app_b2}
\eea
where we have separated the $c_\mr{SW}$ independent part from the part linear
in the clover coefficient.
The precise form of (\ref{app_b1}, \ref{app_b2}) refers to the $U(1)$ gauge
theory; we will include a factor $C_F$ below.


\subsection{Sunset diagram}

With $V^{(n)}_\rh\!=\!\til{h}_{\rh\al}^{(n)}V_\al$ the part of the sunset
diagram proportional to $(ap)^0$ follows from
\beq
[\mr{sunset}]_0/(g_0^2C_F)=\int\!\!{d^4q\ovr(2\pi)^4}\;G(q)
{[V_\rh^{(n)}(q)\!+\!W_\rh^{(n)}]B(q)[V_\rh^{(n)}(q)\!-\!W_\rh^{(n)}]\ovr
\Delta(q)}
\label{app_b3}
\eeq
\bea
V_\rh^{(n)}BV_\rh^{(n)}
&=&f^{2n}V_\rh B V_\rh-
f^n(f^n-1){s_\rh s_\al V_\al B V_\rh+s_\rh s_\be V_\rh B V_\be\ovr s^2}+
(f^n-1)^2{s_\rh^2 s_\al s_\be V_\al B V_\be\ovr(s^2)^2}
\nonumber
\\
&=&f^{2n} V_\rh B V_\rh + (1-f^{2n}){s_\al s_\be\ovr s^2} V_\al B V_\be
\label{app_b4}
\eea
and analogously for $V_\rh^{(n)}BW_\rh^{(n)}$, $W_\rh^{(n)}BV_\rh^{(n)}$
and $W_\rh^{(n)}BW_\rh^{(n)}$.
The terms even in $q$ are
\bea
V_\al B V_\be&\doteq&
2 s^2 s_\al s_\be - 2 \ga_\al \ga_\be c_\al c_\be s^2
+(\ga_\al \ga_\mu c_\al s_\be + \ga_\mu \ga_\be s_\al c_\be ) \bar{s}_\mu
\\
V_\al B W_\be&\doteq&{c_\mr{SW}\ovr2\ri}
(\ga_\al\ga_\mu\si_{\be\la} c_\al c_\be \bar{s}_\mu \bar{s}_\la+
2\si_{\be\la} s^2 c_\be s_\al \bar{s}_\la)
\\
W_\al B V_\be&\doteq&{c_\mr{SW}\ovr2\ri}
(\si_{\al\la}\ga_\mu\ga_\be c_\al c_\be \bar{s}_\mu \bar{s}_\la+
2\si_{\al\la} s^2 c_\al s_\be \bar{s}_\la)
\\
W_\al B W_\be&\doteq&-{c_\mr{SW}^2\ovr2}
\si_{\al\ka}\si_{\be\la} c_\al c_\be s^2 \bar{s}_\ka \bar{s}_\la
\eea
where $\doteq$ stands for ``up to terms odd in $q$''.
With this at hand, we compute
\bea
V_\rh B V_\rh&\doteq&
2(s^2)^2 - 2(4-s^2) s^2 + \bar{s}^2
\\
V_\rh B W_\rh&\doteq&{c_\mr{SW}\ovr2\ri}
(\ga_\rh\ga_\mu\si_{\rh\la} c_\rh^2 \bar{s}_\mu \bar{s}_\la+
\si_{\rh\la} s^2 \bar{s}_\rh \bar{s}_\la)
=\mr{first}+0
\\
W_\rh B V_\rh&\doteq&{c_\mr{SW}\ovr2\ri}
(\si_{\rh\la}\ga_\mu\ga_\rh c_\rh^2 \bar{s}_\mu \bar{s}_\la+
\si_{\rh\la} s^2 \bar{s}_\rh \bar{s}_\la)
=\mr{first}+0
\\
W_\rh B W_\rh&\doteq&-{c_\mr{SW}^2\ovr2}
\si_{\rh\ka}\si_{\rh\la} c_\rh^2 s^2 \bar{s}_\ka \bar{s}_\la
\\
s_\al s_\be V_\al B V_\be&\doteq&
2(s^2)^3 + {1\ovr2} s^2 \bar{s}^2
\\
s_\al s_\be V_\al B W_\be&\doteq&{c_\mr{SW}\ovr2\ri}
({1\ovr4}\ga_\al\ga_\mu\si_{\be\la}\bar{s}_\al\bar{s}_\be\bar{s}_\mu\bar{s}_\la
+\si_{\be\la} (s^2)^2 \bar{s}_\be \bar{s}_\la)
=0
\\
s_\al s_\be W_\al B V_\be&\doteq&{c_\mr{SW}\ovr2\ri}
({1\ovr4}\si_{\al\la}\ga_\mu\ga_\be\bar{s}_\al\bar{s}_\be\bar{s}_\mu\bar{s}_\la
+\si_{\al\la} (s^2)^2 \bar{s}_\al \bar{s}_\la)
=0
\\
s_\al s_\be W_\al B W_\be&\doteq&-{c_\mr{SW}^2\ovr8}
\si_{\al\ka}\si_{\be\la} s^2 \bar{s}_\al \bar{s}_\be \bar{s}_\ka \bar{s}_\la
=0
\eea
where the asserted vanishing of certain terms holds only in case there are no
further factors which destroy the symmetry property it builds on.
With (\ref{app_b4}) we thus arrive at
\bea
V_\rh^{(n)}BV_\rh^{(n)}&\doteq&
f^{2n}(4(s^2)^2-8s^2+\bar{s}^2)+(1-f^{2n})(2(s^2)^2+{1\ovr2}\bar{s}^2)
\nonumber
\\
&=&
2(s^2)^2+{1\ovr2}\bar{s}^2+f^{2n}(2(s^2)^2-8s^2+{1\ovr2}\bar{s}^2)
\\
V_\rh^{(n)}BW_\rh^{(n)}&\doteq&
{c_\mr{SW}\ovr2\ri}f^{2n}
\ga_\rh\ga_\mu\si_{\rh\la}c_\rh^2\bar{s}_\mu\bar{s}_\la
\\
W_\rh^{(n)}BV_\rh^{(n)}&\doteq&
{c_\mr{SW}\ovr2\ri}f^{2n}
\si_{\rh\la}\ga_\mu\ga_\rh c_\rh^2 \bar{s}_\mu \bar{s}_\la
\\
W_\rh^{(n)}BW_\rh^{(n)}&\doteq&
-{c_\mr{SW}^2\ovr2}f^{2n}
\si_{\rh\ka}\si_{\rh\la} c_\rh^2 s^2 \bar{s}_\ka \bar{s}_\la
\eea
making the numerator in (\ref{app_b3}) take the form
\bea
[V_\rh^{(n)}+W_\rh^{(n)}]B[V_\rh^{(n)}-W_\rh^{(n)}]&\doteq&
2(s^2)^2+{1\ovr2}\bar{s}^2+f^{2n}(2(s^2)^2-8s^2+{1\ovr2}\bar{s}^2)
\nonumber
\\
&+&
{c_\mr{SW}\ovr2\ri}f^{2n}
(\si_{\rh\la}\ga_\mu\ga_\rh-\ga_\rh\ga_\mu\si_{\rh\la})
c_\rh^2\bar{s}_\mu\bar{s}_\la
\nonumber
\\
&+&
{c_\mr{SW}^2\ovr2}f^{2n}
\si_{\rh\ka}\si_{\rh\la} c_\rh^2 s^2 \bar{s}_\ka \bar{s}_\la
\;.
\eea
By means of the identities
$\si_{\rh\la}\ga_\mu\ga_\rh-\ga_\rh\ga_\mu\si_{\rh\la}=
2\ri[\ga_\la\ga_\mu-\de_{\la\rh}\de_{\rh_\mu}]$
and
$\si_{\rh\ka}\si_{\rh\la}=\ga_\ka\ga_\la-\ga_\ka\ga_\rh\de_{\rh\la}
-\de_{\ka\rh}\ga_\rh\ga_\la+\de_{\ka\rh}\de_{\rh\la}$,
where in either case $\rh$ is not yet summed over, we thus obtain
\bea
[\mr{sunset}]_0/(g_0^2C_F)&=&{1\ovr2}
\Big[Z_0+\int\!\!{d^4q\ovr(2\pi)^4}\;{f^{2n}\ovr4s^2}\Big]
-2\int\!\!{d^4q\ovr(2\pi)^4}
{f^{2n}\ovr4(s^2)^2+\bar{s}^2}
\nonumber
\\
&+&c_\mr{SW}\int\!\!{d^4q\ovr(2\pi)^4}\;{f^{2n}\ovr4s^2}
{c_\rh^2\bar{s}_\la^2-c_\rh^2\bar{s}_\rh^2\ovr4(s^2)^2+\bar{s}^2}
\nonumber
\\
&+&{c_\mr{SW}^2\ovr2}\int\!\!{d^4q\ovr(2\pi)^4}\;{f^{2n}\ovr4s^2}
{s^2[c_\rh^2\bar{s}_\la^2-c_\rh^2\bar{s}_\rh^2]\ovr4(s^2)^2+\bar{s}^2}
\label{app_b5}
\eea
where $Z_0=\int\!d^4q/(2\pi)^4\,1/(4s^2)=0.15493339...$ has been used.


\subsection{Tadpole diagram}

The tadpole diagram is readily evaluated to give
\bea
[\mr{tadpole}]_0/(g_0^2C_F)&=&
-4\int\!\!{d^4q\ovr(2\pi)^4}\;{G(q)\ovr2}
\sum_\al(\til h_{\rh\al}^{(n)})^2\qquad(\mbox{with $\rh$ fixed})
\nonumber
\\
&=&
-{1\ovr2}\int\!\!{d^4q\ovr(2\pi)^4}\;{1\ovr s^2}
\sum_\al\Big(f^n\de_{\rh\al}-(f^n-1){\qh_\rh\qh_\al\ovr\qh^2}\Big)^2
\nonumber
\\
&=&
-{1\ovr2}\int\!\!{d^4q\ovr(2\pi)^4}\;{1\ovr s^2}
\Big[ f^{2n}-2f^n(f^n-1){\qh_\rh^2\ovr\qh^2}+(f^n-1)^2{\qh_\rh^2\ovr\qh^2}\Big]
\nonumber
\\
&=&
-{1\ovr2}\int\!\!{d^4q\ovr(2\pi)^4}\;{1\ovr s^2}
\Big[ f^{2n} + (1-f^{2n}) {1\ovr d}\Big]
\nonumber
\\
&=&
-{1\ovr2}\Big[Z_0+\int\!\!{d^4q\ovr(2\pi)^4}\;{3f^{2n}\ovr4s^2}\Big]
\label{app_b6}
\eea
where $Z_0=\int\!d^4q/(2\pi)^4\,1/(4s^2)=0.15493339...$ has been used.


\subsection{Combining the two}

It is now straightforward to add (\ref{app_b5}) and (\ref{app_b6}) to obtain
for $am_\mr{crit}\!=\!\Sigma_0$ the result
\bea
-\Sigma_0/(g_0^2C_F)&=&
\int\!\!{d^4q\ovr(2\pi)^4}\;{f^{2n}\ovr4s^2}
+2\int\!\!{d^4q\ovr(2\pi)^4}
{f^{2n}\ovr4(s^2)^2+\bar{s}^2}
\nonumber
\\
&-&c_\mr{SW}\int\!\!{d^4q\ovr(2\pi)^4}\;{f^{2n}\ovr4s^2}\;
{c_\rh^2\bar{s}_\la^2-c_\rh^2\bar{s}_\rh^2\ovr4(s^2)^2+\bar{s}^2}
\nonumber
\\
&-&{c_\mr{SW}^2\ovr8}\int\!\!{d^4q\ovr(2\pi)^4}\;f^{2n}\;
{c_\rh^2\bar{s}_\la^2-c_\rh^2\bar{s}_\rh^2\ovr4(s^2)^2+\bar{s}^2}
\label{app_b7}
\eea
and we comment on the four contributions.
The first term without the $1/(4s^2)$ factor would be
\bea
I^{(m)}&=&\int_{-\pi}^\pi\!\!{dk_1...dk_d\ovr(2\pi)^d}\;
\Big[1-{\al\ovr2(d\!-\!1)}\kh^2\Big]^m
\nonumber
\\
&=&
{d^m\ovr d\si^m}\bigg|_{\si=0}
\int_{-\pi}^\pi\!\!{dk_1...dk_d\ovr(2\pi)^d}\;
e^{\si[1-{\al\ovr2(d\!-\!1)}\kh^2]}
\nonumber
\\
&=&
{d^m\ovr d\si^m}\bigg|_{\si=0}
e^\si\bigg[e^{-{\si\al\ovr d-1}}I_0\Big({\si\al\ovr d-1}\Big)\bigg]^d
\eea
with $m\!=\!2n$, $I_0$ denoting a Bessel function of the second kind and
$I^{(0)}\!=\!1$.
The first term with the denominator but without the smearing factor would
assume the simple form
\bea
J^{(0)}&=&\int_{-\pi}^\pi\!\!{dk_1...dk_d\ovr(2\pi)^d}\;
{1\ovr\kh^2}
\;=\;
\int_0^\infty\!\!d\ta
\int_{-\pi}^\pi\!\!{dk_1...dk_d\ovr(2\pi)^d}\;
e^{-\ta\kh^2}
\nonumber
\\
&=&
\int_0^\infty\!\!d\ta\;
\Big[
e^{-2\ta}I_0(2\ta)
\Big]^d
=Z_0
=\left\{
\begin{array}{cc}
\infty&(d=2)\\
0.25273101...&(d=3)\\
0.15493339...&(d=4)
\end{array}
\right.
\eea
and the actual first contribution can hence be handled via a recursion formula
\bea
J^{(2n)}&=&\int_{-\pi}^\pi\!\!{dk_1...dk_d\ovr(2\pi)^d}\;
\Big[1-{\al\ovr2(d\!-\!1)}\kh^2\Big]^{2n}\,{1\ovr\kh^2}
\nonumber
\\
&=&
\int_{-\pi}^\pi\!\!{dk_1...dk_d\ovr(2\pi)^d}\;
\Big[1-{\al\ovr2(d\!-\!1)}\kh^2\Big]^{2n-1}\,
\Big[{1\ovr\kh^2}-{\al\ovr2(d-1)}\Big]
\nonumber
\\
&=&
J^{(2n-1)}-{\al\ovr2(d-1)}I^{(2n-1)}
\nonumber
\\
&=&
J^{(0)}-{\al\ovr2(d-1)}\Big[I^{(0)}+I^{(1)}+...+I^{(2n-1)}\Big]
\eea
and ditto for $2n\!\to\!m$.
For the other terms we resort to numerical integration.
We collect the pertinent values in Tab.\,\ref{tab_numerics}.
With these it is easy to verify the APE entries in Tab.\,\ref{tab_addshift}.

\begin{table}
\begin{center}
\begin{tabular}{|c|cccc|}
\hline
              & $n\!=\!0$ & $n\!=\!1$ & $n\!=\!2$ & $n\!=\!3$ \\
\hline
$c_\mr{SW}^0$ &  51.43471 &  13.55850 &   7.18428 &   4.81189 \\
$c_\mr{SW}^1$ &  13.73313 &   6.96138 &   4.70457 &   3.56065 \\
$c_\mr{SW}^2$ &  45.72111 &  13.50679 &   6.52280 &   3.84215 \\
\hline
\end{tabular}
\caption{Numerical values of the integrals in (\ref{app_b7}) for
$\al^\mr{APE}\!=\!0.6$ and $n\!=\!0..3$ iterations.}
\label{tab_numerics}
\end{center}
\end{table}


\subsection{Other smearing strategies}

In this article we have focused on a strategy where one applies the same
smearing in three places: in the covariant derivative and the Wilson term of
the Wilson operator (\ref{def_wils}) and in the field-strength tensor of the
clover term (\ref{def_clov}).
Of course, other options are possible.
In general one may apply $n$ steps with parameter $\al$ to build the links
for the (relevant) covariant derivative, $n'$ steps with parameter $\al'$ in
the Wilson term and $n''$ steps with parameter $\al''$ for the clover term.
The numerator in (\ref{app_b3}) then takes the form
$[V_\rh^{(n,n')}(q)\!+\!W_\rh^{(n'')}]B(q)[V_\rh^{(n,n')}(q)\!-\!W_\rh^{(n'')}]$
where $n$ denotes the smearing level in the (relevant) covariant derivative,
$n'$ that in the Wilson term and $n''$ the one in the clover term.
Possible choices include:
\begin{itemize}
\itemsep-1pt
\item
$n=n'=n''=0$: standard (thin-link) clover action (SC)
\item
$n=0, n'=n''>0$: fat-link irrelevant clover action (FLIC), Wilson and clover
terms \cite{Zanotti:2001yb}
\item
$n>0, n'=n''=0$: fat-link relevant clover action (FLRC), only covariant
derivative
\item
$n=n'=n''>0$: fat-link overall clover action (FLOC), same smearing everywhere
\cite{Stephenson:1999ns,DeGrand:2002vu}
\end{itemize}
All explicit numbers given in this article refer to the ``FLOC'' case, but it
is straightforward to generalize the formulae to arbitrary $n,n',n''$.
For instance, for $n=n'$ the terms in (\ref{app_b7}) proportional to
$c_\mr{SW}^0$, $c_\mr{SW}^1$, $c_\mr{SW}^2$ contain a factor $f^{2n'}$,
$f^{n'+n''}$, $f^{2n''}$, respectively.


\section{Details of the parameter dependence}


In this article we focus on the ``standard'' parameters (\ref{std_APE_EXP}) for
APE/EXP smearing and (\ref{std_HYP_HEX}) for HYP/HEX smearing. 
Here, we briefly discuss the dependence on $\al^\mr{APE}\!=\!6\al^\mr{EXP}$.

\begin{table}[!t]
\begin{center}
\begin{tabular}{|c|cccccccc|}
\hline
1\,APE&  0.12  &  0.24  &  0.36  &  0.48  &   0.6  &  0.72  &  0.84  &  0.96  \\
\hline
 $S$  &23.51856&16.57684&11.16131& 7.27195& 4.90876& 4.07175& 4.76092& 6.97626\\
$z_S$ &15.01627&11.34954& 8.30976& 5.89694& 4.11106& 2.95213& 2.42016& 2.51513\\
$z_P$ &17.42350&13.18607& 9.67028& 6.87614& 4.80364& 3.45280& 2.82360& 2.91606\\
$z_V$ &11.74876& 8.75695& 6.35362& 4.53878& 3.31243& 2.67456& 2.62518& 3.16429\\
$z_A$ &10.54515& 7.83869& 5.67337& 4.04918& 2.96614& 2.42423& 2.42346& 2.96382\\
\hline
\end{tabular}
\caption{$S$ and $z_X$ versus smearing parameter for
1\,APE clover fermions with $c_\mr{SW}\!=\!1$.}
\label{tab_pert_alpha_1ape}
\end{center}
\begin{center}
\begin{tabular}{|c|cccccccc|}
\hline
2\,APE&  0.12  &  0.24  &  0.36  &  0.48  &   0.6  &  0.72  &  0.84  &  0.96  \\
\hline
 $S$  &17.57370& 9.36111& 4.99413& 2.77532& 1.66435& 1.27800& 1.89014& 4.43178\\
$z_S$ &11.77989& 6.90750& 3.79645& 1.77701& 0.40606&-0.53293&-1.02988&-0.84811\\
$z_P$ &13.68411& 8.05861& 4.48219& 2.18511& 0.65185&-0.37897&-0.91453&-0.70780\\
$z_V$ & 9.15359& 5.43316& 3.29869& 2.10671& 1.43934& 1.10432& 1.13496& 1.79020\\
$z_A$ & 8.20148& 4.85761& 2.95582& 1.90266& 1.31645& 1.02734& 1.07729& 1.72005\\
\hline
\end{tabular}
\caption{$S$ and $z_X$ versus smearing parameter for
2\,APE clover fermions with $c_\mr{SW}\!=\!1$.}
\label{tab_pert_alpha_2ape}
\end{center}
\begin{center}
\begin{tabular}{|c|cccccccc|}
\hline
3\,APE&  0.12  &  0.24  &  0.36  &  0.48  &   0.6  &  0.72  &  0.84  &  0.96  \\
\hline
 $S$  &13.33394& 5.67072& 2.60949& 1.34010& 0.77096& 0.57512& 1.14061& 4.42535\\
$z_S$ & 9.29121& 4.16584& 1.37936&-0.30406&-1.43930&-2.25393&-2.71777&-2.32827\\
$z_P$ &10.81108& 4.91586& 1.75643&-0.10770&-1.33218&-2.19131&-2.67019&-2.24523\\
$z_V$ & 7.24295& 3.60651& 1.97904& 1.21474& 0.82550& 0.63109& 0.69676& 1.55848\\
$z_A$ & 6.48301& 3.23151& 1.79050& 1.11656& 0.77195& 0.59978& 0.67297& 1.51696\\
\hline
\end{tabular}
\caption{$S$ and $z_X$ versus smearing parameter for
3\,APE clover fermions with $c_\mr{SW}\!=\!1$.}
\label{tab_pert_alpha_3ape}
\end{center}
\end{table}

In Tab.\,\ref{tab_pert_alpha_1ape}-\ref{tab_pert_alpha_3ape} we give details
on how $S$ and $z_X$ for $X\!=\!S,P,V,A$ depend on the smearing parameter
with 1,2,3 steps of APE/EXP filtering with $c_\mr{SW}\!=\!1$.
In most cases, one finds a reduction of $S$ and
$(z_P\!-\!z_S)/2\!=\!z_V\!-\!z_A$ for $\al^\mr{APE}$ between $0$ and
$\sim\!0.75$; beyond that they increase sharply.
This is in line with the discussion in App.\,B -- perturbatively, one expects
larger smearing parameters to be more efficient, up to
$\al^\mr{APE}_\mr{max}\!=\!0.75$ or $\al^\mr{EXP}_\mr{max}\!=\!0.125$.
Hence our ``standard'' choice (\ref{std_APE_EXP}) for the smearing parameter is
not bad -- at least in perturbation theory.

We have also performed a non-perturbative test with $c_\mr{SW}\!=\!1$ clover
fermions on our coarsest lattice, $\be\!=\!5.846$.
We find that $-am_\mr{crit}$ decreases monotonically in the range
$0\!\leq\!\al_\mr{APE}\!\leq\!0.6$.

\begin{table}
\begin{center}
\begin{tabular}{|c|cccccc|}
\hline
$\al^\mr{APE}\!=\!0.6$
      &   0\,APE &  1\,APE & 10\,APE & 100\,APE & 1000\,APE & 10000\,APE  \\
\hline
 $S$  & 31.98644 & 4.90876 & 0.06523 &  0.00063 &  0.00001  & $<\!0.0000001$\\
$z_S$ & 19.30995 & 4.11106 &-5.94036 &-13.18247 & -20.12297 & -27.03399   \\
$z_P$ & 22.38259 & 4.80364 &-5.93562 &-13.18246 & -20.12297 & -27.03399   \\
$z_V$ & 15.32907 & 3.31243 & 0.16719 &  0.01296 &  0.00125  &  0.00013    \\
$z_A$ & 13.79274 & 2.96614 & 0.16482 &  0.01296 &  0.00125  &  0.00013    \\
\hline
\end{tabular}
\caption{$S$ and $z_X$ versus iteration number for $\al^\mr{APE}\!=\!0.6$
clover fermions with $c_\mr{SW}\!=\!1$.}
\label{tab_pert_iter}
\end{center}
\end{table}

With $n_\mr{iter}\to\infty$ one expects in perturbation theory that $S$ and
$(z_P\!-\!z_S)/2\!=\!z_V\!-\!z_A$ tend to zero.
We checked this explicitly, with details given in Tab.\,\ref{tab_pert_iter}.
The approach seems to be monotonic in $n_\mr{iter}$; we do not observe any
oscillations.

\clearpage





\end{document}